\documentclass{pasj00}
\newcommand{\asca}{{\it ASCA} }

\newcommand{\rosat}{{\it ROSAT} }
\newcommand{\xmm}{{\it XMM-Newton} }

\newcommand{\chandra}{{\it Chandra} }
\newcommand{\chandrap}{{\it Chandra}}

\newcommand{\suzaku}{{\it Suzaku} }
\newcommand{\suzakup}{{\it Suzaku}}

\newcommand{\rxte}{{\it RXTE} }

\newcommand{\fekalfa}{{Fe~K$\alpha$} }
\newcommand{\fekalfap}{{Fe~K$\alpha$}}
\newcommand{\fekbeta}{{Fe~K$\beta$} }
\newcommand{\fekbetap}{{Fe~K$\beta$}}
\newcommand{\ekalfa}{{$E_{{\rm Fe~K}\alpha}$} }
\newcommand{\ekalfap}{{$E_{{\rm Fe~K}\alpha}$}}
\newcommand{\ikalfa}{{$I_{{\rm Fe~K}\alpha}$} }
\newcommand{\ikalfap}{{$I_{{\rm Fe~K}\alpha}$}}
\newcommand{\ekbeta}{{$E_{{\rm Fe~K}\beta}$} }
\newcommand{\ekbetap}{{$E_{{\rm Fe~K}\beta}$}}
\newcommand{\ikbeta}{{$I_{{\rm Fe~K}\beta}$} }
\newcommand{\ikbetap}{{$I_{{\rm Fe~K}\beta}$}}

\newcommand{\feklya}{{Fe~{\sc xxvi}~Ly$\alpha$} }

\newcommand{\feklyap}{{Fe~{\sc xxvi}~Ly$\alpha$}}

\newcommand{\bsax}{{\it BeppoSAX} }

\newcommand{\figprelima}{{figure~1(a)} }
\newcommand{\figprelimap}{{figure~1(a)}}

\newcommand{\figprelimbp}{{figure~1(b)}}
\newcommand{\figdatspecrat}{{figure~2} }

\newcommand{\figfekprof}{{figure~3(a)} }

\newcommand{\figfekprofc}{{Figure~3(a)} }

\newcommand{\figbroadprof}{{figure~3(b)} }
\newcommand{\figbroadprofp}{{figure~3(b)}}
\newcommand{\figewvsfwhm}{{figure~3(c)} }
\newcommand{\figewvsfwhmp}{{figure~3(c)}}
\newcommand{\figidvsin}{{figure~3(d)} }
\newcommand{\figidvsinp}{{figure~3(d)}}
\newcommand{\figrvstheta}{{figure~5} }
\newcommand{\figrvsthetap}{{figure~5}}

\newcommand{\figekbvseka}{figure~4(a) }
\newcommand{\figekbvsekap}{figure~4(a)}
\newcommand{\figikbvseka}{figure~4(b) }
\newcommand{\figikbvsekap}{figure~4(b)}
\newcommand{\figfeion}{figure~6 }

\newcommand{\tableobslogp}{table~1}

\newcommand{\tableresultsp}{table~2}
\newcommand{\tableresultsc}{Table~2 }

\newcommand{\apec}{{\tt APEC} }

\begin{document}
\SetRunningHead{T. Yaqoob et al.}{Precision \fekalfa and \fekbeta Spectroscopy of NGC~2992}
\Received{2006 August 2}%{yyyy/mm/dd}
\Accepted{2006 September 14}%{yyyy/mm/dd}

\title{Precision \fekalfa and \fekbeta Line 
Spectroscopy of the Seyfert 1.9 Galaxy NGC~2992 with {\it Suzaku}}

%%% begin:list of authors
\author{Tahir \textsc{Yaqoob},\altaffilmark{1,2}
Kendrah D. \textsc{Murphy},\altaffilmark{1,2}
Richard E. \textsc{Griffiths},\altaffilmark{3}
Yoshito \textsc{Haba},\altaffilmark{4} \\
Hajime \textsc{Inoue},\altaffilmark{5}
Takeshi \textsc{Itoh},\altaffilmark{4}
Richard \textsc{Kelley},\altaffilmark{2}
Motohide \textsc{Kokubun},\altaffilmark{6} \\
Alex  \textsc{Markowitz},\altaffilmark{2,7} 
Richard \textsc{Mushotzky},\altaffilmark{2} 
Takashi \textsc{Okajima},\altaffilmark{1,2} 
Andrew \textsc{Ptak},\altaffilmark{1,2} \\
James \textsc{Reeves},\altaffilmark{1,2} 
Peter J. \textsc{Serlemitsos},\altaffilmark{2} 
Tadayuki \textsc{Takahashi},\altaffilmark{5} 
and
Yuichi \textsc{Terashima}\altaffilmark{5,8}}
%  \thanks{Example: Present Address is xxxxxxxxxx}}
%\affil{A-Address of Institute}
%\email{aaaaa@xxx.xxx.xx.xx}

%\author{B-Firstname \textsc{B-Familyname}}
%\affil{B-Address of Institute}\email{bbbbb@xxx.xxx.xx.xx}
%\and
%\author{C-Firstname {\sc C-Familyname}}
%\affil{C-Address of Institute}\email{ccccc@xxx.xxx.xx.xx}
%%% end:list of authors

%%% Please use the following style in case that sorting by 
%%% affilation is impossible. 
%
% \author{%
%   D-Firstname \textsc{D-Familyname}\altaffilmark{1}
%   E-Firstname \textsc{E-Familyname}\altaffilmark{1,2}
%   and
%   F-Firstname \textsc{F-Familyname}\altaffilmark{2}}
% \altaffiltext{1}{Address of Institute}
% \email{ddddd@xxx.xxx.xx.xx}
% \email{eeeee@xxx.xxx.xx.xx}
% \altaffiltext{2}{Address of Institute}

\altaffiltext{1}{Department of Physics and Astronomy, 
Johns Hopkins University, \\ 3400 N. Charles St., Baltimore, MD 21218}
%\email{yaqoob@pha.jhu.edu}
\altaffiltext{2}{Exploration of the Universe Division,
NASA Goddard Space Flight Center, \\ Greenbelt Rd., Greenbelt, MD 20771}
\altaffiltext{3}{Department of Physics, Carnegie Mellon University, 
5000 Forbes Avenue, \\ Pittsburgh, PA 15213}
\altaffiltext{4}{Department of Astrophysics, School of Science,  Nagoya University,
Chikusa-ku, \\ Nagoya 464-01, Japan}
\altaffiltext{5}{Institute of Space and Astronautical Science, \\ Japan
Aerospace Exploration Agency, 3-1-1 Yoshino-dai, \\
Sagamihara, Kanagawa 229-8510, Japan}
\altaffiltext{6}{Department of Physics, University of Tokyo, 
7-3-1 Hongo, Bunkyo-ku, \\ Tokyo, Japan}
\altaffiltext{7}{NASA Postdoctoral Research Associate}
\altaffiltext{8}{Present address: Department of Physics, Ehime University, \\
Bunkyo-cho, Matsuyama, Ehime 790-8577, Japan}

%% `\KeyWords{}' always has to be placed before `\maketitle'.
 %Do NOT move this preamble from here!

%\vspace{10cm}
\maketitle

\begin{abstract}
We present detailed time-averaged X-ray spectroscopy in the 0.5--10~keV
band of the Seyfert~1.9 galaxy NGC~2992 with the \suzaku
X-ray Imaging Spectrometers (XIS). 
The source had a factor $\sim 3$ higher 2--10~keV flux 
($\sim 1.2 \times 10^{-11} \rm \ erg \ cm^{-2} \ s^{-1}$) than the 
historical minimum and a factor $\sim 7$ less than the historical maximum.
The XIS spectrum of NGC~2992 can be described by several components.
There is a primary continuum, modeled as a power-law 
with a photon index of $\Gamma = 1.57^{+0.06}_{-0.03}$ that is
obscured by a Compton-thin absorber
with a column density of $8.0^{+0.6}_{-0.5} \times \rm 10^{21} \ cm^{-2}$.
There is another, weaker, unabsorbed power-law component 
(modeled with the same slope as the primary),
that is likely to be due to the primary continuum being 
electron-scattered into our line-of-sight by
a region extended on a scale of hundreds of parsecs.
We measure the
Thomson depth of the scattering zone 
to be $\tau_{\rm es} = (0.073 \pm 0.021)/[\Omega/4\pi]$,
where $\Omega/4\pi$ is the fraction of the sky covered
by the zone (as seen from the X-ray source) that is visible to the observer.
An optically-thin thermal emission component, which probably
originates in the same extended region, is included in the model
and yields a temperature and luminosity of $kT=0.656^{+0.088}_{-0.061}$~keV
and $\sim 1.2 \pm 0.4 \times 10^{40} \rm \ erg \ s^{-1}$ respectively.
We detect an Fe~K emission complex which we model with broad and narrow
lines and we show that the intensities of the two components are decoupled
at a confidence level $>3\sigma$. The broad \fekalfa line has an
equivalent width of $118^{+32}_{-61}$~eV and could originate in
an accretion disk (with inclination angle greater than $\sim 30^{\circ}$)
around the putative central black hole. The narrow \fekalfa line has an 
equivalent width of $163^{+47}_{-26}$~eV and is unresolved 
(FWHM~$<4090 \rm \ km \ s^{-1}$)
and likely originates in distant matter.
The absolute flux
in the narrow line implies that the column density out of the
line-of-sight could be much higher than measured in the line-of-sight, and
that the mean (historically-averaged) continuum luminosity responsible for forming the line
could be a factor of several higher than that measured from the data.
We also detect the \fekbeta line 
(corresponding to the narrow \fekalfa line)
with a high signal-to-noise ratio
and describe a new robust
method to constrain the ionization state of Fe responsible
for the \fekalfa and \fekbeta lines that does not require
any knowledge of possible gravitational and Doppler energy shifts
affecting the line energies. 
For the distant line-emitting matter (e.g. the putative obscuring
torus) we deduce that the predominant ionization state is 
lower than Fe~{\sc viii} (at 99\% confidence), 
conservatively taking into account 
residual calibration
uncertainties in the XIS energy scale and theoretical and experimental
uncertainties  in the Fe~K fluorescent line energies.
From the limits on a possible Compton-reflection continuum
it is likely that the narrow \fekalfa and \fekbeta lines originate in
a Compton-thin structure. 
\end{abstract}

\section{Introduction}
\label{intro}

The Fe~K emission line profiles in active galactic nuclei (AGN)
generally consist of a prominent core centered at $\sim 6.4$~keV
(typically with FWHM~$<10,000  \rm \ km \ s^{-1}$ or so) and
in some sources an additional underlying broader component.
The relatively narrow core at $\sim 6.4$~keV is found to be 
ubiquitous in nearby, low to moderate luminosity AGN (e.g. Nandra \etal~1997a,b;
Sulentic \etal~1998; Weaver, Gelbord, \& Yaqoob 2001; Reeves 2003;
Page \etal~2004; Yaqoob \& Padmanabhan 2004;
Jim\'{e}nez-Bail\'{o}n \etal~2005; 
Zhou \& Wang 2005; Jiang, Wang, \& Wang 2006; Levenson \etal~2006).
The broad Fe~K line component potentially carries
signatures of the inner accretion disk in the vicinity of the putative
black hole by virtue of
gravitational and Doppler energy shifts
(e.g. Tanaka \etal~1995; Nandra \etal~1997a,b;
Reynolds 1997; Turner \etal~1998; Iwasawa \etal~1999; Yaqoob \etal~2002;
Reynolds \& Nowak 2003; Fabian \& Miniutti 2005).
The bulk of the Fe~K line core is likely to originate in matter located
much further from the central black hole than the outer disk
(e.g. the outer parts of the optical broad line 
region, the putative parsec-scale torus structure, or the optical narrow line region).
In addition to the two principal components,
there are an increasing number of reports of cases of narrow lines at energies higher
than 6.4~keV (from ionized and/or outflowing
Fe -- see Bianchi \etal~2005, and references therein) 
and lower than 6.4~keV (from apparently redshifted
localized emission -- Turner \etal~2002; Iwasawa, Miniutti, \& Fabian 2004;
Turner \etal~2006, and references therein).

Deconvolving the disk and distant-matter components of the
Fe~K line is extremely challenging because the disk can also
contribute to the line core emission (from radii greater than $\sim 
100r_{g} \equiv 100GM/c^{2}$ or so). 
In other words, line emission from disks observed at low inclination
angles (relative to the disk normal) is partially degenerate with
line emission from distant matter.
Line emission in the wings of
the profile from regions close to the black hole is smeared out
over energy and therefore difficult to detect against the continuum.
In order to derive the most reliable parameters and constraints
for the disk and distant-matter Fe~K line components, 
{\it one must simultaneously model both}, otherwise the model
parameters will be biased and misleading (e.g. see 
discussions in Weaver \& Reynolds 1998,
Yaqoob \& Padmanabhan 2004). So far only some of the brightest AGN
have sufficient signal-to-noise ratio to produce useful
constraints from dual-line models (e.g. see Yaqoob \etal~2002,
Yaqoob \& Padmanabhan 2004, and references
therein). For the brightest sources the Fe~K line core is
usually unresolved even with the \chandra gratings because 
high spectral resolution {\it and} high
throughput are required.
Therefore, the location of the ``distant'' Fe~K line emitter remains unknown.
Nevertheless, decoupling the intensities of the broad and
narrow lines (in the sense of showing that the data require both
to be non-zero at some confidence level) would be an important step
in constraining models of the central engine in AGN. 
Such decoupling has only been possible for a few of the brightest
AGN (e.g. see Fabian \& Miniutti 2005). 
We also note that detection of the broad Fe~K line in type~1.5--2 AGN
is even more difficult than in type~1 objects because the X-ray spectra
of the former class are more complex and are usually affected by 
significant intrinsic X-ray absorption 
(e.g. see Turner \etal~1998; Weaver \& Reynolds 1998;
Matt \etal~2000; Guainazzi \etal~2001; Risaliti 2002). 

Almost all of our knowledge of the Fe~K line emission in AGN comes
from the \fekalfa line. Although \fekbeta line emission has been 
reported in a few type~1.5--2 AGN (Sambruna \etal~2001;
Molendi, Bianchi, \& Matt~2003; Matt \etal~2004), constraints on the line energy and 
equivalent width (EW) have been loose because the branching ratio,
\fekbetap/\fekalfap, is less than 0.145 (Palmeri \etal~2003, and
references therein). Yet the \fekbeta line carries
{\it additional} information on the ionization state of line-emitting Fe that
would supplement the information carried by the \fekalfa line.

In this paper we present the results of detailed X-ray CCD spectroscopy
of the nearby ($z=0.00771$, Keel 1996) Seyfert~1.9 galaxy
NGC~2992 using data from \suzaku observations.
We show that the intensities of the broad and narrow \fekalfa lines
are decoupled for the first time in this source. We also detected the
\fekbeta line corresponding to the \fekalfa line core and present measurements
of its centroid energy, intensity, and EW. We describe a new method
that utilizes the \fekbeta line to deduce tight limits on the ionization
state of Fe in the matter responsible for emitting the line core.
For the \fekalfa and \fekbeta line analysis we modeled the 0.5--10~keV
spectrum self-consistently so we also report the broadband results.

NGC~2992 has been observed by every X-ray astronomy mission 
since the time it was discovered by HEAO-1 to be one of the brightest
hard X-ray AGN in the sky, with a 2--10~keV flux of
$\sim 7.2-8.6 \times
10^{-11} \rm \ erg \ cm^{-2} \ s^{-1}$ (Piccinotti \etal~1982). 
In more than a quarter of a century of X-ray observations, the hard X-ray
flux has varied by over a factor of 20, corresponding to a
range in the intrinsic 2--10~keV luminosity of $\sim 0.55-11.8
\times 10^{42} \rm \ erg \ s^{-1}$ (assuming $H_{0} = 70 \
\rm km \ s^{-1} \ Mpc^{-1}$, $\Lambda = 0.73$, $\Omega = 1$).
During an \asca
observation in 1994, NGC~2992 was in its lowest continuum flux
state thus far observed, with a 2--10~keV flux of
$\sim 4 \times 10^{-12} \rm \ erg \ cm^{-2} \ s^{-1}$ (Weaver \etal~1996).
During that observation the Fe K line equivalent width (EW) was
very high ($\sim 500-700$~eV), indicating that
the line intensity had not responded to the declining continuum.
In 1997 a \bsax observation found NGC~2992 to be still in a
low continuum state, and the Fe~K line EW was even higher, but
then a second \bsax observation in 1998 revealed that the source
continuum had fully recovered to its bright HEAO-1 state (see Gilli \etal~2000).
In this state, the EW
of the Fe~K line was very low ($<100$~eV) since the line intensity
again did not respond to the change in continuum level. These measurements
were mostly sensitive 
to the Fe~K line core, which from the lack of variability,
likely arises in a distant, parsec-scale reprocessor, as
suggested by Weaver \etal~(1996). 

The results from an \xmm observation of NGC~2992 in May 2003 are unpublished.
We examined those data and found that they suffered
from heavy pile-up. A detailed analysis of the \xmm data is beyond the
scope of this work and will be reported elsewhere.

NGC~2992 was the subject of an \rxte monitoring
campaign consisting of 24 observations
in the period 2005 March to 2006 January. The \suzaku observations
that are the subject of the present paper 
were made during the \rxte campaign and were quasi-simultaneous with some of the
\rxte observations. The scope of the present paper is restricted
to the \suzaku data only, but a detailed analysis of the data from the \rxte campaign
will be presented elsewhere.

The present paper is organized as follows.
In section~\ref{data} we describe the observations and details of the data reduction.
In section~\ref{spfit} we describe the spectral fitting analysis and in section~\ref{results} we
discuss the detailed results of that analysis, including
brief comparisons with historical data and results. In section~\ref{kakbanal} we
describe a new method to tightly constrain the ionization state of Fe using
the narrow \fekalfa and \fekbeta emissions lines. In section~\ref{concl} we 
summarize our results and give our conclusions.

\section{Observations and Data Reduction}
\label{data}
The joint Japan/US X-ray astronomy satellite, \suzaku (Mitsuda \etal~2006), 
was launched on 10 July, 2005. 
NGC~2992 was observed on three
occasions in 2005, November and December (the observation log is given
in \tableobslogp).
\suzaku carries four X-ray Imaging Spectrometers (XIS -- Koyama \etal~2006) and 
a collimated Hard X-ray Detector (HXD -- Takahashi \etal~2006).
Each XIS consists of four CCD detectors
at the focal plane of its own thin-foil X-ray telescope
(XRT -- Serlemitsos \etal~2006), and has a field-of-view (FOV) of $17.8' \times 17.8'$.
One of the XIS detectors (XIS1) is
back-side illuminated (BI) and the other three (XIS0, XIS2, and XIS3) are
front-side illuminated (FI). The bandpass of the FI detectors is
$\sim 0.4-12$~keV and $\sim 0.2-12$~keV for the BI detector. The useful
bandpass depends on the signal-to-noise ratio of the source since
the effective area is significantly diminished at the extreme ends of the
operational bandpasses. Although the BI CCD has higher effective area
at low energies, the background level across the entire bandpass is higher
compared to the FI CCDs. 
The spectral resolution of the XIS detectors has been continuously
degrading with time but the current instrument response
matrices employ the resolution before the onset of degradation, 
as described in Koyama \etal~(2006).
Although we used these standard response matrices for modeling the
XIS data,
we measured the widths of the  Mn~$K\alpha$ lines from the on-board
$^{55} \rm Fe$ calibration sources (two per XIS) using the actual
NGC~2992 observations in order to correctly interpret spectral
fitting results. Details are given in section~\ref{meanxis}.

\begin{table*}[!htb]
\caption{NGC 2992 {\it Suzaku} observation log.
$^{a}$ Start and end times correspond to the time tags of the first and last
photons respectively in the cleaned and filtered events files combined from XIS2 and XIS3.
$^{b}$ Mean (0.5--10~keV)
count rates and exposure times are for XIS2 and XIS3 {\it per XIS}.
}
 \begin{center}
\begin{tabular}{lcccccc}
Obs & Start$^{a}$ & & End$^{a}$ & & Count Rate$^{b}$ & Exposure \\
& (UT) & & (UT) & & (ct/s/XIS) & Time$^{b}$ (s) \\
& & & & & & \\
1 & 6/11/2005 & 14:16:51 & 7/11/2005 & 14:01:50 & 0.3888 $\pm$ 0.0027 & 34664.5 \\
2 & 19/11/2005 & 21:42:59 & 20/11/2005 & 23:24:45 & 0.5096 $\pm$ 0.0032 & 31641.5 \\
3 & 13/12/2005 & 10:15:37 & 14/12/2005 & 12:08:38 & 0.4408 $\pm$ 0.0026 & 41681.0 \\
\end{tabular}
\end{center}
\end{table*}

\subsection{HXD Data}

The HXD consists of  
two non-imaging instruments (the PIN and GSO -- see Takahashi \etal~2006)
with a combined bandpass of $\sim 10-600$~keV.
Both of the HXD instruments are background-limited. We found that for the
NGC~2992 observations the average on-source count rate in the full HXD/PIN
band ($\sim 10-120$~keV) was $\sim 0.75$ ct/s, whilst the corresponding estimated
background count rate was 0.65 ct/s so that NGC~2992 constitutes only $\sim 13\%$
of the total signal. The situation is even worse for the GSO which has a
smaller effective area than the PIN.
In order to obtain reliable background-subtracted spectra,
the background spectrum must be modeled as a function of energy and time.
The background model is still under development and cannot yet reproduce
the background spectral and temporal behavior with sufficient accuracy 
to give reliable HXD spectra for NGC~2992. Therefore, in this paper
we will restrict our analysis to the XIS only and defer analysis of the
simultaneous HXD data to future work when the background model systematics are
lower. The observations of NGC~2992 were in fact optimized for the HXD
in terms of positioning the source at the aimpoint for the HXD (the so-called
``HXD-nominal pointing'') which gives a somewhat lower count-rate in the XIS
than the ``XIS-nominal'' pointing but gives $\sim 10\%$ higher HXD effective area.

\subsection{XIS Data Filtering, Cleaning, and Selection}
\label{clnxis}

The XIS observations of NGC~2992 were made in so-called $3\times 3$
and $5 \times 5$ edit modes which correspond to 9 and 25 pixel pulse-height
arrays respectively that are telemetered to the ground (see Koyama \etal~2006
for details).
The array is centered on the pixel containing the largest
pulse height of an event. We combined the data from both modes as
there did not appear to be any noticeable differences in the spectra 
obtained from the different modes.
Data from version 0.7 of the pipeline reprocessing 
(Fujimoto \etal~2006; Mitsuda \etal~2006) were used for the analysis 
in the present paper. As is standard practice, only {\it ASCA} grades
0, 2, 3, 4, and 6 were selected. Event files were cleaned 
to remove hot and
flickering pixels with the
FTOOL {\tt cleansis} as updated for the XIS. Data were
excluded during satellite passages through the South Atlantic Anomaly
(SAA), and for time intervals less than $256$~s after passages through the SAA,
using the T\_SAA\_HXD house-keeping parameter.
Data were also rejected for Earth elevation angles (ELV) less than $5^{\circ}$,
Earth day-time elevation angles (DYE\_ELV) less than $20^{\circ}$, and values
of the magnetic cut-off rigidity (COR) less than 6 ${\rm GeV}/c^{2}$.
Version 0.7 of the reprocessing pipeline also includes CTI corrections
based on available in-flight calibration and other refinements to
the XIS energy scale. Residual uncertainties in the energy scale
are on the order of 0.2\% or less
(or $\sim 13$~eV at 6.4~keV -- see Koyama \etal~2006). We confirmed this
from an analysis of the onboard calibration line data taken during the
NGC~2992 observations (see section~\ref{meanxis} for details). 

The cleaning and data selection resulted in net exposure times 
in the range 31.6~ks to 41.7~ks for the three observations (see \tableobslogp).
We examined XIS images made from the cleaned events files and found that
the galaxy NGC~2993 ($\sim 3'$ from the core of NGC~2992) was clearly detected
but its total count rate was less than 1\% of the NGC~2992 count rate.
Even though the half-power diameter of the XRT$+$XIS is $\sim 2'$,
NGC~2992 and NGC~2993 are resolved. NGC~2993 was detected
by \chandra with much better spatial resolution (see Colbert \etal~2005).
We extracted XIS lightcurves and spectra using
two sets of circular regions
with radii of 2.6' and 4.35' centered on NGC~2992. The smaller region
excludes NGC~2993, resulting in negligible contamination of the NGC~2992
spectra,
and the larger region corresponds to the size assumed by some of the
standard XRT effective area files that have been released.
The spectra from the smaller regions will be used for the principal analysis
and those from the larger regions will be used to better estimate
absolute fluxes (see below).
We extracted background spectra using circular regions
with radii of $\sim 4'$ from source-free regions of the detectors. 

\subsection{Instrument Response Files and the Mean XIS Spectrum}
\label{meanxis}

For spectral analysis we used the response matrices (or ``RMF'' files) 
{\tt ae\_xi0\_20060213.rmf}, {\tt ae\_xi1\_20060213c.rmf},
{\tt ae\_xi2\_20060213.rmf}, and {\tt ae\_xi3\_20060213.rmf} for 
XIS0, XIS1, XIS2 and
XIS3 respectively (binned to the same number of channels
as the spectra). The telescope effective area files (or ``ARF'' files)
used were {\tt ae\_xi0\_hxdnom4\_20060415.arf}, {\tt ae\_xi1\_hxdnom4\_20060415.arf},
{\tt ae\_xi2\_hxdnom4\_20060415.arf} and {\tt ae\_xi3\_hxdnom4\_20060415.arf}
for XIS0, XIS1, XIS2, and XIS3 respectively. 
The ARF files
are standard files released for the HXD-nominal pointing position and
assume a circular extraction with a radius of 2.9'. Since our source
extraction regions had a radius of 2.6' we derived a mean correction
factor by comparing broadband fluxes measured from the spectra
extracted using 4.35' radii regions with those from our 2.6'
radius region. The ARF files used for the spectra from the larger
regions were {\tt ae\_xi0\_hxdnom6\_20060415.arf}, 
{\tt ae\_xi1\_hxdnom6\_20060415.arf}, {\tt ae\_xi2\_hxdnom6\_20060415.arf} 
and {\tt ae\_xi3\_hxdnom6\_20060415.arf} for XIS0, XIS1, XIS2, and XIS3
respectively. 
Although the larger regions include NGC~2993,
the contamination is small and the procedure is adequate for the present
purpose. 

Some preliminary examination and spectral fitting revealed
no significant evidence for variability in the total
count rate or spectral shape within each of the three observations. Although
there was up to $\sim 30\%$ variability in net count rate between observations,
we found no statistically significant evidence for spectral variability
between the observations (see \tableobslogp). 
We found that the four XIS sensors were consistent with each other
to $\sim 10\%$ or better (within statistical errors) over the
energy band $0.5-10$~keV.
However, XIS0 showed a systematic dip in the $\sim 5-7$~keV
band (i.e. in the region that \fekalfa line emission is expected) which
was not apparent in the other three XIS sensors.
Also, XIS1 showed a broad dip in the $\sim 2-5$~keV band which was
not apparent in the other three XIS sensors. Although both of these
effects are at the level of $\sim 10\%$ or less, for the sake of
maximum conservatism,
in the remainder of this paper all of the
spectral analysis will refer to a single XIS spectrum combined from observations
1, 2 and 3, using only XIS2 and XIS3 data. 
For analysis of the mean spectrum from
XIS2$+$XIS3 (integrated over the three observations), we combined 
the RMF and ARF files (four in total)
into one response file with appropriate weightings, according
to the exposure times of the data segments. 
The mean 0.5--10~keV background-subtracted count rate for this spectrum was 
$0.4442 \pm 0.0016$ ct/s {\it per XIS}, and the
net exposure time was $1.08 \times 10^{5}$~s {\it per XIS}. 
We restricted the analysis to the 0.5--10~keV
band in order to avoid background-subtraction systematics
(in the 0.5--10~keV band the background constitutes $\sim 3.4\%$ of the
total count rate).
We obtained a flux correction factor for the
size of the extraction region of 2.1\% (as described above) for the mean
spectrum. All fluxes and
absolute line intensities quoted in this paper have been corrected (increased)
by this factor.

We measured the centroid energies
of the Mn~$K\alpha$ lines from the calibration sources,
using a background-subtracted spectrum combined from XIS2 and XIS3 data, summed over
observations 1--3 (i.e. the same combination used for the
analysis of NGC~2992), and summed over both calibration sources
on each XIS. The expected energies of the Mn~$K\alpha_{1}$ and
Mn~$K\alpha_{2}$ lines are 5.89875~keV and 5.88765~keV respectively
(Bearden 1967). Since the $K\alpha_{1}:K\alpha_{2}$ branching
ratio is 2:1, the expected centroid energy is then 5.89505~keV.
Using a single Gaussian we measured a centroid energy of 
$5.906 \pm 0.003$~keV (one-parameter, 90\% confidence error).
Therefore, the difference between the measured and expected 
centroid energy is $11 \pm 3$~eV, consistent with the
accuracy of the absolute XIS energy scale given by Koyama \etal~(2006)
as 0.2\% ($\sim 13$~eV) at the \fekalfa line energy.
Both the
residual CTI correction uncertainty and the spectral resolution are
better at the centers of the detectors (i.e. in the region of the aimpoints
for source observations) than at the detector edges where the calibration
sources are located so this is an over-estimate. 
We also measured the Mn~$K\alpha$ line width
and obtained a FWHM of 161.5~eV (at an energy of
$\sim 5.9$~keV).  At 6.4~keV
this corresponds to a FWHM of 168.4~eV (see Koyama \etal~2006),
or $\sim 7890 \rm \ km \ s^{-1}$ FWHM. Since the FWHM 
resolution at 6.4~keV assumed in the response matrices is
133.8~eV (from direct inspection), and 
measured line widths are a convolution of the 
assumed response with input lines, there is a residual FWHM width
of $\sqrt{(168.4^{2} - 133.8^{2})} = 102.3$~eV FWHM 
(or $4795 \rm \ km \ s^{-1}$ at 6.4~keV)
that is not currently accounted for in the instrument response function.
Again, the spectral resolution is better at the center of the detectors
(by $\sim 5-10$~eV) than at the edges.

\subsection{XIS Contaminant Correction}
\label{contam}

The XIS detectors have suffered a loss in low-energy efficiency due to
absorption by a contaminant on the optical blocking filter. The degradation
can be modeled using ``effective'' column densities of Carbon and Oxygen
that can be calculated as a function of time since launch and position on each XIS,
based on empirical fits to in-flight data obtained from non-varying astrophysical
sources (see Koyama \etal~2006). We calculated the Carbon column density for each XIS and
each observation of NGC~2992. For analysis of the mean spectrum described
above we then calculated the average of the six column densities for XIS2 and XIS3,
observations 1 to 3, and obtained $3.4 \times 10^{18} \rm \ cm^{-2}$.
Based on the current assumption that the contaminant is probably 
$\rm C_{24} H_{38} O_{4}$, the O column density was fixed at 1/6 of the value
of the C column density.
The C and O column densities were then combined with photoelectric absorption
cross-sections from Balucinska-Church \& McCammon (1992) since 
these were used to derive the empirical degradation formulae. The resulting
contamination model, implemented using the {\tt varabs}
model in XSPEC, was included along with all astrophysical models of NGC~2992
(see section~\ref{spfit}).

\section{Spectral Fitting}
\label{spfit}

In the remainder of this paper we shall refer to analysis of the
combined spectrum from XIS2 and XIS3 averaged over all three observations (\tableobslogp),
as described in section~\ref{meanxis}. 
We used XSPEC v11.3.2 for spectral fitting (Arnaud 1996), utilizing
data in the 0.5--10~keV band. The instrument response files used were
described in section~\ref{meanxis}. 
The spectrum was binned to $\sim 29$~eV per bin and this resulted
in $>25$ ct/bin for the net spectrum over the 0.5--9.7~keV range so
we were able to use the $\chi^{2}$ statistic for finding the best-fitting
model and for the statistical error analysis. 
Unless otherwise stated, all statistical errors given correspond to
90\% confidence for one interesting parameter ($\Delta\chi^{2} = 2.706$).
Galactic absorption with a  column density of $5.26 \times 10^{20} \ \rm cm^{-2}$ 
(Dickey \& Lockman 1990) was included in all of the models
described hereafter and its inclusion will be implicitly assumed.
For the Galactic absorption, as well as any intrinsic absorption, we used
photoelectric cross-sections given by Morrison \& McCammon (1983).
The correction for the XIS contamination degradation
(as described in section~\ref{contam}) was also included in all
of the spectral fitting described below.
All astrophysical
model parameter values will be given in the rest frame of NGC~2992.

\subsection{Preliminary Spectral Fitting and the Baseline Model}
\label{baseline}

%Figures:
%Figure 1
\begin{figure}
  \begin{center}
     \rotatebox{0}{\FigureFile(80mm,112mm){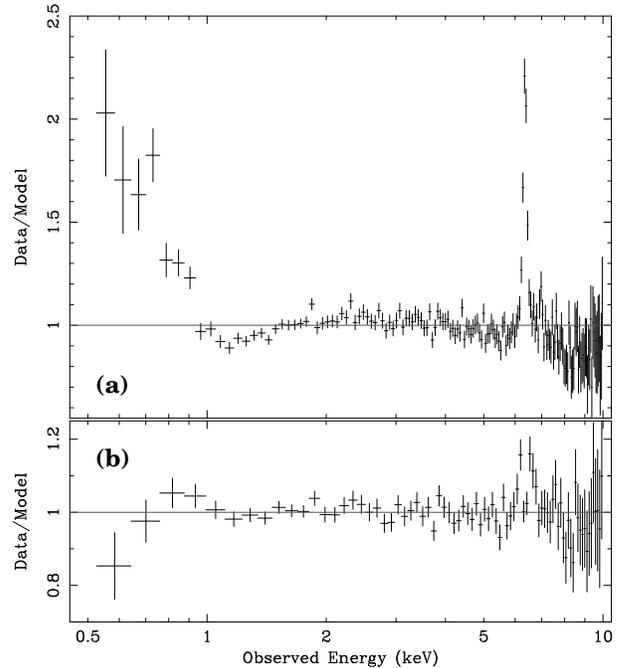}}
  \end{center}
  \caption{(a) Ratio of the XIS2$+$XIS3 spectrum (integrated over observations 1 to 3) to
a model consisting of Galactic absorption, an intrinsic power law continuum,
and a uniform intrinsic absorber. (b) As (a), but now
the model has additional components corresponding to an unabsorbed power-law
component (for example due to scattering), and narrow (unresolved) Gaussian
\fekalfa and \fekbeta
emission-line components. See section~\ref{baseline} for details.
}
\end{figure}

Preliminary spectral fitting confirmed the complexity in the X-ray spectrum
of NGC~2992 that has been established from observations with previous
X-ray astronomy missions (see section~\ref{intro}).
This complexity is illustrated in \figprelimap, which shows the ratio of the
XIS data to a simple model consisting of a power-law intrinsic continuum and
uniform intrinsic absorption fully covering the source. There are only three
free parameters in this model, namely the normalization of the 
power-law continuum, its photon index, $\Gamma$, and the column density, $N_{H}$.
For this 
model $\chi^{2} =1158.2$ for 321 degrees of freedom.
Relative to this simple model there is a large excess at soft X-ray energies
below $\sim 2$~keV and a very prominent \fekalfa emission line. In addition,
a weaker emission line is apparent at $\sim 7$~keV, which could be due to
\feklyap, \fekbetap, or a blend of both. From a detailed analysis described
in section~\ref{kakbanal} we identify the weaker line with \fekbeta and 
show quantitatively that contamination by any \feklya line emission is negligible.

At least part of the soft excess in \figprelima is due to scattering of the intrinsic 
X-ray continuum into the line-of-sight by an extended optically-thin zone
in the active nucleus (e.g. Turner \etal~1998). 
This scattered component has been imaged 
in NGC~2992 using \rosat and \chandra (see Colbert \etal~2005,
and references therein) and in some other
Seyfert 1.5--2 galaxies (e.g. Morse \etal~1995; Ogle \etal~2000;
Young, Wilson, \& Shopbell 2001; George \etal~2002; Iwasawa \etal~2003). 
The direct continuum along the line-of-sight
is obscured so even if only a small fraction of the intrinsic continuum is
scattered, the soft X-ray spectrum can easily be dominated by the scattered
component. Therefore, in the next step we added another power-law continuum
component to the model, with the photon index tied to that of the direct
continuum and a normalization that was allowed to float. We parameterized
this normalization in terms of the fraction, $f_{s}$, of the directly observed
power-law continuum normalization. In the optically-thin limit the Thomson
depth, $\tau_{\rm es}$,
of the scattering region (assuming spherical symmetry) is simply equal to
$f_{s}$. More realistically, $f_{s} = [\Omega/4\pi] \tau_{\rm es}$,
where $[\Omega/4\pi]$ is the solid angle as a fraction of $4\pi$
subtended by the scattering zone at the X-ray source. Only that
part of the scattering zone visible to the observer should be
included in this solid angle. In NGC~2992 an extended
bi-conical outflow has been imaged (e.g. Allen \etal 1999; Veilleux
\etal 2001), with opening angles of 
$\sim 125^{\circ}-135^{\circ}$. Taking this angle to be $130^{\circ}$, we get 
$[\Omega/4\pi] \sim 1-\cos{75^{\circ}}$ for the two cones together,
or $f_{s} \sim 0.74 \tau_{\rm es}$.

We also added two unresolved Gaussian components to
model the emission lines in the Fe~K band. The intrinsic widths ($\sigma_{N}$)
of the two lines were tied together 
(in this case the two emission
lines originate from the same atoms/ions). In the preliminary
spectral fitting $\sigma_{N}$ was fixed at a value (0.005~keV) much less than the 
XIS spectral
resolution. The other parameters were the centroid energies of the two emission
lines (\ekalfa and \ekbetap), and their intensities (\ikalfa and \ikbeta
respectively).
Thus, in this more complex model there were a total of eight free parameters
($\Gamma$, $N_{H}$, $f_{s}$, \ekalfap, \ekbetap, \ikalfap, \ikbetap,
and a continuum normalization). 
The best-fitting model gave $\chi^{2} =400.6$ for 316 degrees of freedom.
The ratio of the XIS data to this best-fitting model is shown in \figprelimbp.
The fit is still poor and statistically significant residuals are still 
evident in the soft X-ray band
below $\sim 2$~keV and in the Fe~K band. Note that if we included
only one Gaussian component (to model the strongest emission line)
we obtained $\chi^{2} =418.8$ for 318 degrees of freedom, demonstrating
that the weaker emission line is statistically significant (see section~\ref{kakbanal}
for a more quantitative analysis).
We then allowed the intrinsic width ($\sigma_{N}$) of the Gaussian emission-line components
to be free and re-fitted the model, obtaining another significant reduction in
$\chi^{2}$, which dropped to 375.4 for 
315 degrees of freedom. 

Motivated by the fact that an extended soft thermal emission component has been
detected in NGC~2992 (see Colbert \etal~2005) and in other type~1.5--2
Seyfert galaxies (e.g. Ptak \etal~1997; Bianchi, Guainazzi, \&
Chiaberge~2006, and
references therein), we then
added a component to model optically-thin thermal emission. We used the
\apec model in XSPEC with the element abundances fixed at
cosmic values (but later
we allowed the Fe abundance to be free in order to derive constraints
on it). The \apec
model then involves two more free parameters, $kT$, and $L_{\rm APEC}$,
the temperature and luminosity of the plasma respectively 
(although we used the normalization
of the \apec model 
as the actual free parameter in the fits, not its luminosity).
In all, there were now eleven free parameters and we obtained 
$\chi^{2} =354.1$ for 313 degrees of freedom. 

There still remained some residuals in the Fe~K band indicative of additional,
broader, line emission. This might be expected from an X-ray illuminated 
accretion disk whose inner radius extends to within $\sim 10$ or so
gravitational radii of the putative central black hole. Such relativistically
broadened Fe~K line emission has been observed in other AGN 
(e.g. Fabian \& Miniutti 2005, and references therein). 
Indeed, it is possible that the Gaussian emission-line components in the above
preliminary model could be modeling part of the disk line emission as well
as the narrow core emission that might originate from more distant matter.
With CCD data the line core can be degenerate with part of the disk line component
because both components can have significant emission near the line rest-frame
energy. 
Even though the Fe~K line core may be the dominant component, it is
in general 
necessary to simultaneously model both the broad and narrow components 
otherwise the inferred model parameters
for one component will be biased by the presence of
the unmodeled component. For example, in the spectral fits described above
with the Gaussian widths free, the best-fitting lines may be broader
than the true width of the Fe~K line core because they may be trying
to model an underlying broad component as well.

Therefore, we added Fe~K line emission 
from a relativistic disk around a (Schwarzschild) black hole 
using the {\tt diskline} model in XSPEC (see Fabian \etal~1989).
The {\tt diskline} model has six parameters, namely the energy of the line in
the disk frame, $E_{0}$; the inner and outer radii
of the disk 
($R_{\rm in}$ and $R_{\rm out}$ respectively,
in units of gravitational radii, $r_{g} \equiv GM/c^{2}$);
the power-law index, $q$, of the radial line emissivity law ($r^{q}$);
the inclination angle of the disk normal 
relative to the observer's line-of-sight,
$\theta_{\rm obs}$; and the integrated intensity of the line, $I_{\rm disk}$. 
The overall Fe~K line profile in the XIS data is already dominated by
the line core and we found that the data for any residual broad-line
emission could not constrain all of the disk line model parameters.
In particular, there is considerable degeneracy between $E_{0}$ and $\theta_{\rm obs}$.
Therefore, we fixed $E_{0}$ at 6.40~keV (corresponding to the
rest-frame energy of Fe~{\sc i}~K$\alpha$). In addition, we fixed
$R_{\rm in}$ and $R_{\rm out}$ at $6r_{g}$ 
(the radius of the last stable orbit around a Schwarzschild 
black hole) and $1000r_{g}$ respectively.
For self-consistency we also included a disk line component corresponding
to Fe~{\sc i}~K$\beta$ with {\it no} additional free parameters:
$R_{\rm in}$, $R_{\rm out}$, $q$, and $\theta_{\rm obs}$ were forced
to take the same values as the \fekalfa line,  
$E_{0}$ was fixed at 7.056~keV, and the intensity of the 
\fekbeta line was forced to be 13.5\% of the \fekalfa line
intensity (see section~\ref{kakbanal} for a detailed discussion
on \fekbeta line energies and the Fe~K$\beta$/K$\alpha$ branching ratio). 
Although there may be no clear isolated detection
of the {\it broad} \fekbeta line, for ionization states of Fe~{\sc i} to 
Fe~{\sc xvi} it {\it will be produced}, and excluding it from the model could affect
the values of the inferred parameters of the other Fe~K line components.
Fitting the XIS data with the Fe~K disk line emission components
included in the model results in  a drop in $\chi^{2}$ of 19.7
(to 334.4 for 310 degrees of freedom) for an extra three free parameters
($q$, $\theta_{\rm obs}$, and $I_{\rm disk}$), indicating that a broad-line
component is required by the data with a confidence level of more than
$3\sigma$. 

%Figure 2
\begin{figure}
  \begin{center}
    \rotatebox{0}{\FigureFile(80mm,112mm){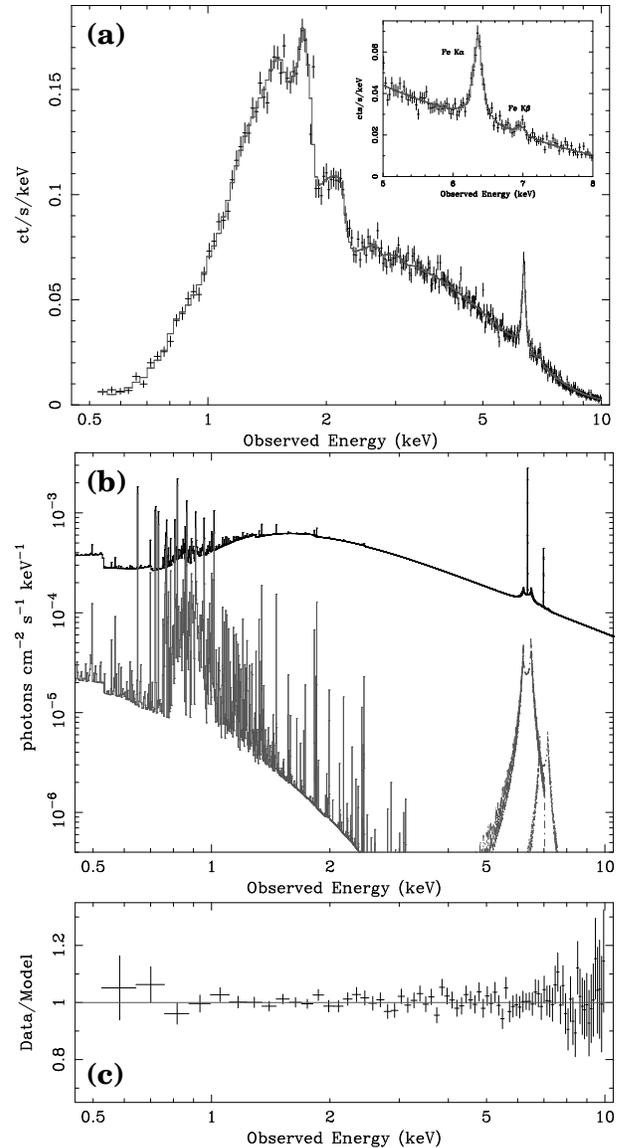}}
  \end{center}
  \caption{(a) The XIS2$+$XIS3 counts spectrum (integrated over observations 1 to 3)
overlaid with the best-fitting baseline model (solid line). The inset shows a close-up of
the region containing the \fekalfa and \fekbeta emission lines.
The C/O XIS contamination is included in the model.
(b) The best-fitting baseline model (see section~\ref{baseline} and
\tableresultsp). The soft X-ray
optically-thin thermal continuum and the broad
\fekalfa and \fekbeta disk-line components are also shown separately.
The C/O XIS contamination model has been removed in the plot but Galactic
absorption is included.
(c) The ratio of the data in (a) to the best-fitting baseline model.
}
\end{figure}

Finally, we added a component to model a Compton-reflection continuum
using the model {\tt hrefl} in XSPEC (see Dov\v{c}iak, Karas, \& Yaqoob 2004).
Although only the $\sim 7-10$~keV band XIS data are able to constrain
the reflection continuum, if any of the Fe~K emission line components
originate in Compton-thick matter, a reflection continuum {\it will be
produced}. Excluding a reflection continuum in the model could bias
the best-fitting parameters of the Fe~K emission lines.
The reflection model involves only one extra free parameter, $R$, which
is the normalization of the reflection continuum relative to the
case of a steady-state X-ray continuum illuminating a slab subtending
a solid angle of $2\pi$ at the X-ray source. The inclination angle between the
disk normal and the observer's line-of-sight was tied to the disk inclination
angle for the relativistic Fe~K emission-line components. Note that a
disk geometry is likely to be inappropriate for any 
reflection continuum associated with Compton-thick distant-matter
components of the Fe~K line emission, but the data are not
sensitive to differences in the reflection continuum due to geometry. 
When we fitted the above model to the data we found a best-fitting
value of $R=0$, and consequently no improvement in the fit. 
Nevertheless, we retained the reflection continuum in all of the spectral
fitting analysis (with $R$ free),
since it could potentially affect the statistical error
analysis for the other parameters.

Thus, our baseline model consisted of fifteen free parameters:
two normalizations (one for the soft, optically-thin thermal continuum,
another for
the intrinsic power-law continuum), $kT$, $\Gamma$, $N_{H}$, $f_{s}$,
$\sigma_{N}$, \ekalfap, \ikalfap, \ekbetap, \ikbetap, $q$, $\theta_{\rm obs}$,
$I_{\rm disk}$, and $R$. This model gave an excellent fit to the data
($\chi^{2} = 334.4$ for 309 degrees of freedom), as
can be seen from \figdatspecrat(a) (which shows the best-fitting model overlaid
on the XIS counts spectrum), \figdatspecrat(b) (which shows the best-fitting model
and some of the individual model components), and \figdatspecrat(c)
(which shows the ratio of the XIS data to the best-fitting model).
The best-fitting parameters and 90\%, one-parameter statistical errors are
shown in \tableresultsp. We discuss these results in detail in section~\ref{results}.

\begin{table}[!htb]
\caption{XIS Spectral fitting results for NGC~2992.
All values and energy ranges in this table (except $F_{\rm 0.5-2 \ keV}$
and $F_{\rm 2-10 \ keV}$) refer to the rest-frame of NGC~2992
($z=0.00771$). Statistical errors are for 90\%, one interesting parameter
($\Delta \chi^{2} = 2.706$) and were derived with fifteen parameters
free, with $I_{\rm Ni \ Fe \ K\alpha}$ fixed at 0. Upper limits on the
intensity and equivalent width (EW) of the Ni~K$\alpha$ line were derived
with sixteen parameters free. \\
$^{a}$ The upper limits on the
widths of the line core of Fe~K$\alpha$ and Fe~K$\beta$ have {\it not} been corrected
for
the degradation of the XIS spectral resolution (see section~3.2).
$^{b}$ Upper limits on $R$ are for $\theta_{\rm obs}$ fixed at the
best-fitting value of $43^{\circ}$ and for $\theta_{\rm obs}=60^{\circ}$,
as indicated. $^{c}$ Observed-frame fluxes, corrected for the
CCD contamination but {\it not} corrected for Galactic and intrinsic absorption.
$^{d}$ Intrinsic, rest-frame luminosities, corrected for all
absorption components.
}
 \begin{center}
\begin{tabular}{lr}

$\chi^{2}$ / degrees of freedom & 334.4/309 \\
& \\
$kT$ (keV) & $0.658^{+0.088}_{-0.061}$ \\
$L_{\rm APEC}$ ($10^{40} \rm \ erg \ s^{-1}$) & $1.18^{+0.36}_{-0.45}$ \\
& \\
$\Gamma$ & $1.569^{+0.056}_{-0.027}$ \\
$N_{H} \rm \ (10^{21} \ cm^{-2})$ &  $7.99^{+0.56}_{-0.45}$ \\
$f_{\rm s}$ (scattered fraction) & $0.073^{+0.021}_{-0.021}$ \\
& \\
$\theta_{\rm obs}$ (degrees) & $>31$ \\
$q$  & $-1.5_{-0.8}^{+5.1}$ \\
$I_{\rm disk} \rm \ [Fe~K\alpha]$ ($\rm 10^{-5} \ photons \ cm^{-2} \ s^{-1}$)
& $1.9^{+0.5}_{-1.0}$ \\
$\rm EW_{disk} \ [Fe~K\alpha]$ (eV) & $118^{+32}_{-61}$ \\
& \\
$E_{N} \rm \ [Fe~K\alpha]$ (keV) & $6.407^{+0.007}_{-0.007}$ \\
$\sigma_{N}$ (keV) $^{a}$ & $<0.044$ \\
FWHM ($\rm km \ s^{-1}$) $^{a}$ & $<4850$ \\
$I_{\rm N} \rm \ [Fe~K\alpha]$ ($\rm 10^{-5} \ photons \ cm^{-2} \ s^{-1}$)  &
$2.49^{+0.71}_{-0.40}$ \\
$\rm EW_{N} \ [Fe~K\alpha]$ (eV)  & $163^{+47}_{-26}$ \\
$E_{N} \ [\rm Fe~K\beta]$ (keV) & $7.034^{+0.037}_{-0.038}$ \\
$I_{\rm N} \rm \ [Fe~K\beta]$ ($\rm 10^{-5} \ photons \ cm^{-2} \ s^{-1}$) &
$0.43^{+0.20}_{-0.23}$ \\
$\rm EW_{N} \ [Fe~K\beta]$ (eV) & $27^{+12}_{-14}$ \\
& \\
$I_{\rm N} \rm \ [Ni~K\alpha]$ ($\rm 10^{-5} \ photons \ cm^{-2} \ s^{-1}$) & $<0.29$ \\
 $\rm EW_{N} \ [Ni~K\alpha]$ (eV) & $<28$ \\
& \\
Reflection fraction, $^{b}$ $R (\theta_{\rm obs} = 43^{\circ})$ & $<0.60$ \\
Reflection fraction, $^{b}$ $R (\theta_{\rm obs} = 60^{\circ})$ & $<0.78$ \\
& \\

$F_{\rm 0.5-2 \ keV}$ ($10^{-11} \rm \ erg \ cm^{-2} \ s^{-1}$)$^{c}$ & 0.16 \\
$F_{\rm 2-10 \ keV}$ ($10^{-11} \rm \ erg \ cm^{-2} \ s^{-1}$)$^{c}$  & 1.15 \\
$L_{\rm 0.5-2 \ keV}$ ($10^{42} \rm \ erg \ s^{-1}$)$^{d}$ & 0.70 \\
$L_{\rm 2-10 \ keV}$ ($10^{42} \rm \ erg \ s^{-1}$)$^{d}$  & 1.60 \\

\end{tabular}
\end{center}

\end{table}

\subsection{Deconvolution of the Broad and Narrow Fe~K Emission Lines}
\label{deconv}

\begin{figure*}
  \begin{center}
    \rotatebox{0}{\FigureFile(134mm,234mm){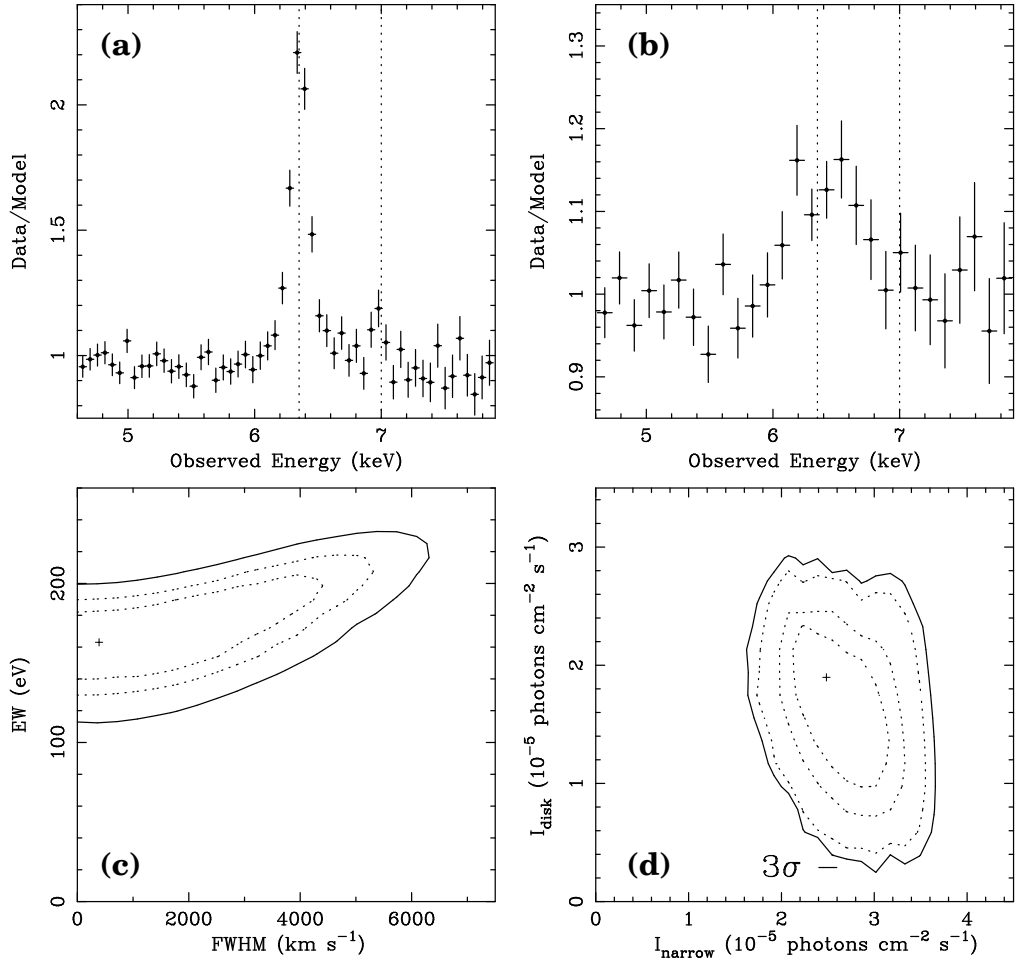}}
  \end{center}
  \caption{(a) A close-up of the XIS spectral region in \figprelima containing
the \fekalfa and \fekbeta lines. The vertical dotted lines correspond to
the observed-frame energies of Fe~{\sc i} K$\alpha$ and K$\beta$ at
6.400~keV and 7.056~keV in the rest-frame respectively.
All of the spectral analysis in the present paper refers to
a mean spectrum constructed from XIS2$+$XIS3, integrated over observations 1 to 3.
(b) The ratio of the XIS spectral data to the baseline model
re-fitted after removing the relativistic \fekalfa and \fekbeta disk-line components,
whilst freezing the optically-thin thermal continuum, narrow Fe~K line, and
Compton-reflection continuum
parameters at their best-fitting values (see section~\ref{deconv} for details).
(c) The 68\%, 90\%, 99\% confidence contours of the equivalent width
(EW) versus the intrinsic width (FWHM) of the narrow, distant-matter \fekalfa line
based on the best-fitting baseline model (section~\ref{baseline}).
The contours have {\it not} been corrected for the degradation in XIS
spectral resolution. After correction the 99\% confidence, two-parameter
upper limit is $\sim 4090 \rm \ km \ s^{-1}$ instead of
$\sim 6300 \rm \ km \ s^{-1}$ (see section~\ref{deconv}).
(d) The 68\%, 90\%, 99\%, and
$3\sigma$ confidence contours of the \fekalfa relativistic disk-line intensity
($I_{\rm disk}$)
versus the intensity of the narrow, distant-matter \fekalfa line
($I_{\rm narrow} \equiv I_{N}$)
based on the best-fitting baseline model (section~\ref{baseline}).
It can be seen that even at a confidence
level of $3\sigma$, the disk and distant-matter \fekalfa lines are
decoupled in the sense that the intensity of either of the two line
components is non-zero everywhere on the $3\sigma$ contour.
}
\end{figure*}

\figfekprofc shows a close-up of the Fe~K region of the XIS spectrum of
NGC~2992 in the form of the ratio of the data to a model consisting of a simple power-law
continuum with a fully-covering absorber (i.e. \figfekprof is simply a ``zoomed'
region from \figprelimap). Note that the \fekbeta line is clearly seen here.
A broad base to the prominent \fekalfa line core is also apparent. In
section~\ref{baseline} we found that when we modeled the underlying broad emission
with a disk line, the Gaussian components that modeled the line core emission
became narrower than for the case when the broad line emission was not modeled.
In fact the best-fitting baseline model (\tableresultsp) 
indicates that the line cores of \fekalfa and \fekbeta are unresolved by the XIS
data. In order to better illustrate the Fe~K line profile due to the
putative relativistic disk emission, we removed the disk line model components
from the baseline model, froze the parameters of the thermal emission,
reflection continuum, and Gaussian line core models and re-fitted the data.
The resulting ratio of data to model is shown in \figbroadprof which 
illustrates the
profile of the relativistic disk component of the Fe~K line. 

Next, starting from the baseline model again, with its fifteen free parameters
(section~\ref{baseline}) we constructed joint confidence contours of the 
equivalent width (EW) of the \fekalfa line core versus its intrinsic width
(FWHM). The 68\%, 90\%, 99\% confidence contours are shown in \figewvsfwhm
which demonstrates that the 99\%, two-parameter upper limit on the FWHM
appears to be 
$\sim 6300 \ \rm km \ s^{-1}$. 
However, as mentioned in section~\ref{data} the response matrices
do not include the time-dependent degradation in the XIS energy
resolution. In section~\ref{meanxis} we showed that analysis of the
Mn~$K\alpha$ calibration line data taken during the NGC~2992
observations gave a residual FWHM width
of 102.3~eV (or $4795 \rm \ km \ s^{-1}$ at 6.4~keV)
that is not accounted for by the current response matrices.
Since the observed line width from any observed source is a
convolution of the intrinsic line width and the instrumental
broadening, the true 99\% confidence,
two-parameter upper limit of the \fekalfa line
core in NGC~2992 is then $[6300^{2} - 4795^{2}]^{\frac{1}{2}}  
\sim 4090 \rm \ km \ s^{-1}$. Note that the narrow-line widths
and their one-parameter statistical errors 
in \tableresultsp, the relativistic disk line parameters,
and the contours in \figewvsfwhm
have {\it not} been corrected for this effect.
The one-parameter upper limit on the narrow-line width is similar
to the residual error in the width of the response function
($\sim 102$~eV FWHM or $\sigma \sim 43$~eV).
The disk line is measured to be a little broader than it
really is
so that our lower limit on the disk inclination angle is
slightly over-estimated and the radial emissivity parameter, $q$,
should be slightly smaller.

We also constructed joint confidence contours
of the 
intensity of the Fe~K$\alpha$ disk line versus the Fe~K$\alpha$ intensity of the 
narrow, distant-matter line. During this process, as for \figewvsfwhmp,
{\it all of the thirteen
remaining parameters in the baseline model were free}. 
The resulting contours are shown in \figidvsinp, for 68\%, 90\%, 99\%, and
$3\sigma$ respectively. Remarkably, it can be seen that at a confidence
level of greater than $3\sigma$, the disk and distant-matter \fekalfa lines are
decoupled in the sense that the intensity of either of the two line 
components is non-zero everywhere on the $3\sigma$ contour.

We investigated the possibility that part of the disk line may be
an artifact due to the current XIS response matrices inadequately
modeling the low-energy tail of the response function 
(see Koyama \etal~2006). Again we used the Mn~$K\alpha$ calibration line 
background-subtracted spectrum, integrated
over observations 1--3, summed over XIS2 and XIS3, using both calibration
sources on each XIS. We modeled the main peak with a disk line and Gaussian
line, plus a simple power-law continuum. The disk-line model 
parameters (see section~\ref{baseline}) $E_{0}$, $R_{\rm in}$,
$R_{\rm out}$ were fixed at the values
used for modeling NGC~2992 and the values of
$\theta_{\rm obs}$, and $q$
were fixed at the best-fitting values shown in \tableresultsp.
The disk line and Gaussian line intensities, as well as the peak energy and width of
the Gaussian component were free parameters. 
By constructing confidence contours of the disk line intensity versus the
Gaussian line intensity we established that the low-energy
excess in the actual in-flight
XIS response can only account for a disk line intensity
that is $2\%$ of the Gaussian line intensity, with a 99\% confidence 
two-parameter
upper limit of $\sim 6\%$. Yet for the actual NGC~2992 data we
obtained a best-fitting \fekalfa disk line intensity that was 76\% of the line core
(see \tableresultsp). Even the $3\sigma$ confidence contour of the disk line
intensity versus the line core intensity (see \figidvsinp) only goes down to
a value of $\sim 8\%$ for the ratio of the disk line to line core intensity.
Therefore the low-energy excess in the tail of the XIS response does
not have a significant impact on the measurements of the disk line in the
NGC~2992 data or its decoupling from the \fekalfa line core.

Decoupling of the \fekalfa components from the accretion disk
and distant matter has not been possible with previous data for NGC~2992 
(see section~\ref{fekresults}).
It is even more remarkable that the decoupling has been
possible with CCD resolution data and for a relativistic disk line that
is not particularly impressive in terms of signal-to-noise ratio
(\figbroadprofp) or EW (see \tableresultsp).
The reason can be traced to the fact that the {\it narrow line} has
excellent signal-to-noise and a large EW. This in turn leaves less 
freedom for the disk line model.
While it is possible that there may not be two distinct physical
locations of \fekalfa line emission, a single line-emitting region
would demand a peculiar line-emissivity law as a function of radius.
The strength of the line core compared to the broad base (which we
model as a disk line) would demand that the emissivity fall as
a function of radius and then increase again. The width of the
broad base of the \fekalfa line (see \figbroadprofp) is $\sim 33,000
\ \rm km \ s^{-1}$~FWHM, more than five times the upper limit
on the width
of the unresolved line core (\figewvsfwhmp). Yet a uniform spherical
distribution of matter gives line profiles which have much
less contrast in the line width between the center and wings of the
profile (e.g. see Yaqoob \etal~1993).
The implications of the parameters derived from the dual \fekalfa line
modeling will be discussed further in section~\ref{fekresults} and
compared with published results on historical data.

\section{Discussion of Results}
\label{results}

In this section we discuss in detail the results that
were obtained by fitting the
baseline model (section~\ref{baseline}) to the XIS data (see \tableresultsp).

\subsection{Soft, Optically-thin Thermal Emission}
\label{apecresults}

We obtained a temperature and luminosity of the optically-thin thermal
emission component of $kT=0.656^{+0.088}_{-0.061}$~keV and
$\sim 1.2 \pm 0.4 \times 10^{40} \rm \ erg \ s^{-1}$ respectively.
The temperature is typical of that found for a similar component in other
type~1.5--2 AGN (e.g. see Ptak \etal~1997; Bianchi \etal~2006,
and references therein). Note that the errors quoted here on the temperature
are statistical only and do not include systematic errors, which may be
larger. \chandra was able to image
the soft X-ray emission better than \suzaku but was still not able
to spatially separate it from other emission components due to 
contamination from the PSF wings 
caused by the dominant nuclear emission (Colbert \etal~2005).
Compounded with this were CCD pile-up issues, and lower signal-to-noise ratio
of the data. Therefore \chandra obtained somewhat worse constraints on the temperature
of the soft thermal emission. Colbert \etal~(2005) give $kT = 0.51^{+0.26}_{-0.19}$~keV
and $L = 1.4 \times 10^{40} \rm \ erg \ s^{-1}$ and conclude that the
bulk of the soft X-ray emission originates from within $\sim 150-500$~pc of
the central engine. The temperature and luminosity of the extended soft
emission component obtained by \chandra are both consistent with our 
respective values measured with \suzakup.

Our \suzaku measurement of the temperature of the soft thermal emission
in NGC~2992 is the most reliable to date.
Elvis \etal~(1990) reported detecting the soft X-ray emission component in
NGC~2992 using the {\it Einstein} HRI, but it was only detected
in one out of four quadrants, the temperature was poorly
constrained, and the luminosity was more than an order of magnitude greater than
that obtained by \suzaku and \chandrap. Although the soft X-ray emission
imaged by the \rosat HRI was 
spatially consistent with \chandra (see Colbert \etal~2005),
the \rosat HRI could not measure the temperature since it was not a 
spectroscopic instrument. The \rosat PSPC spectral resolution was 
too poor to measure the temperature. \asca CCD spectra had better spectral
resolution but the signal-to-noise ratio of the NGC~2992 \asca data is
lower than that of the \suzaku data (see Weaver \etal~1996). 
{\it BeppoSAX} could not constrain the soft thermal component
because both the MECS and LECS instruments had poorer spectral
resolution than the \suzaku XIS  and the useful MECS bandpass did not extend to
below $\sim 0.9$~keV. 

\subsection{Intrinsic Continuum and Absorption}
\label{abscont}

We obtained $\Gamma = 1.57^{+0.06}_{-0.03}$ for the power-law
photon index of the intrinsic continuum (see \tableresultsp). This is
somewhat less than that obtained from \bsax ($\Gamma \sim 1.7$ -- Gilli \etal~2000)
and \chandra ($\Gamma \sim 1.8$ -- Colbert \etal~2005) observations.
\asca could not constrain $\Gamma$ for the complex models required to
explain the data (Weaver \etal~1996).
Given the model-dependence of $\Gamma$ and residual instrument
cross-calibration uncertainties of the order of $\Delta \Gamma \sim 0.1$,
we do not interpret the apparent differences
in $\Gamma$ amongst the different
observations as necessarily having an astrophysical origin.

We derived a column density of $8.0^{+0.6}_{-0.5} \times  10^{21} \rm \ cm^{-2}$
for the line-of-sight X-ray obscuration 
(see \tableresultsp) which corresponds to a
Thomson depth of $\sim 0.006^{+0.005}_{-0.004}$ (i.e. the absorber
is Compton-thin, at least in the line-of-sight). This column density is
consistent with values obtained from previous 
observations with \bsax (Gilli \etal~2000) and \asca (Weaver \etal~1996).

We tested the data to see if any additional Fe~K absorption was required
by adding a simple absorption-edge model 
({\tt zedge} in XSPEC) with the threshold energy fixed at 7.11~keV
(corresponding to neutral Fe) but allowing the optical depth at the
threshold energy to be a free parameter. All the other 15 free parameters
in the model remained free. We obtained a best-fitting threshold optical
depth of zero and a one-parameter, 90\% confidence upper limit of 0.06.
Since the optical depth at the Fe~K edge threshold energy for the
absorption already included in the model is $\sim 0.008$ (for $N_{H} =
8 \times  10^{21}  \rm \ cm^{-2}$ and the Fe abundance used), we 
could only obtain
a loose upper limit on the possible over-abundance of Fe of $\sim 7.5$.

\tableresultsc gives the {\it observed} XIS fluxes (including the effect of
Galactic and intrinsic absorption but not the CCD contaminant)
in the 0.5--2~keV and 2--10~keV bands which
were 0.16 and $1.15 \times 10^{-11} \rm \ erg \ cm^{-2} \ s^{-1}$ 
respectively. However, note that systematic uncertainties
in the absolute flux calibration of X-ray detectors, as indicated by
flux reference
values quoted for the Crab Nebula observed by different instruments,
are generally a few percent but can be as high as 10-20\% in some cases
(e.g. see Jahoda \etal~2006, and references therein). 
Historically, the lowest and highest 2--10~keV observed fluxes
for NGC~2992 have been measured by \asca in 1994 ($\sim 0.4
\times 10^{-11} \rm \ erg \ cm^{-2} \ s^{-1}$ -- Weaver \etal~1996), and
by {\it HEAO-1} in 1978 
($\sim 8.6 \times 10^{-11} \rm \ erg \ cm^{-2} \ s^{-1}$
-- see Piccinotti \etal~1982) respectively. This represents a factor
$\sim 21.5$ in dynamic range. The \suzaku flux is at the lower end of
the historically observed range, but it is nearly a factor $\sim 3$ higher
than the historical minimum. 

\tableresultsc also gives the inferred XIS luminosities
{\it after correcting for absorption}. These were calculated
by setting the Galactic and intrinsic absorbing column densities
to zero in the best-fitting baseline model (section~\ref{baseline}). We obtained
0.7 and $1.6 \times 10^{42} \rm \ erg \ s^{-1}$ in the
0.5--2~keV and 2--10~keV (rest-frame) bands respectively (assuming $H_{0} = 70 \
\rm km \ s^{-1} \ Mpc$, $\Lambda = 0.73$, $\Omega = 1$).
The historical range in 2--10~keV luminosity is therefore $\sim 0.55-11.8
\times 10^{42} \rm \ erg \ s^{-1}$. 

\subsection{Scattered Continuum}
\label{scatcont}

\tableresultsc shows that we measured $(7.3 \pm 2.1)\%$ for the continuum scattered into the
line-of-sight (as a percentage of the  direct line-of-sight continuum
before absorption).
This translates directly into $[\Omega/4\pi] \tau_{\rm es} = 0.073 \pm 0.021$,
where $[\Omega/4\pi]$ is the fraction of the sky covered by
the extended, warm, electron scattering zone, as seen from the X-ray
source, including only that part of the scattering zone visible to the observer.
Using $[\Omega/4\pi] \sim 0.74$ for the bi-conical outflow in NGC~2992
(see \S\ref{baseline}), we get $ \tau_{\rm es} = 0.099 \pm 0.028$.
The scattered fraction and optical depth cannot be easily compared with
results from previous observations. The \asca data had very poor
signal-to-noise and the model in Weaver \etal~1996 did not include
extended thermal emission. The \bsax data did not have coverage of the
soft X-ray band with sufficient resolution to measure $f_{s}$
(see Gilli \etal~2000).

Colbert \etal~(2005) determined from \chandra ACIS data that the
scattering zone must be situated at more than $\sim 150$~pc from the nucleus
since the spectrum from inside this region could be fitted with a
power-law and absorption only. However, despite \chandra having much
better spatial resolution than \suzakup, the \chandra data could not
constrain the optical depth of the electron-scattering zone 
because the scattered continuum is spatially confused with the soft thermal
emission and even with some of the direct central emission, due to the
PSF wings. Spectroscopically, the \chandra ACIS data are inferior
to the \suzaku data because they suffer from severe pile-up and have a much
lower signal-to-noise ratio than the \suzaku data. For example, the
0.3--8~keV nuclear,
and ``near-nuclear'' \chandra spectra analyzed by Colbert \etal~(2005)
had $\sim 6,744$ and $\sim 14,600$ counts respectively, as opposed to 
a total of nearly $\sim 10^5$ counts (0.5--10~keV) in the XIS spectrum 
that we have analyzed here. 
Note also that pile-up in the \chandra data prohibited a
reliable measurement of the flux and luminosity of the nuclear source.
Thus, even though the spatial resolution
of \suzaku is not as good as \chandrap, the high signal-to-noise
ratio in the \suzaku spectrum and absence of 
pile-up problems allowed a better measurement of the
scattered component and hence a better estimate of
the Thomson depth of the scattering zone.

\subsection{Fe~K Emission Lines}
\label{fekresults}

\begin{figure}
  \begin{center}
    \rotatebox{0}{\FigureFile(80mm,140mm){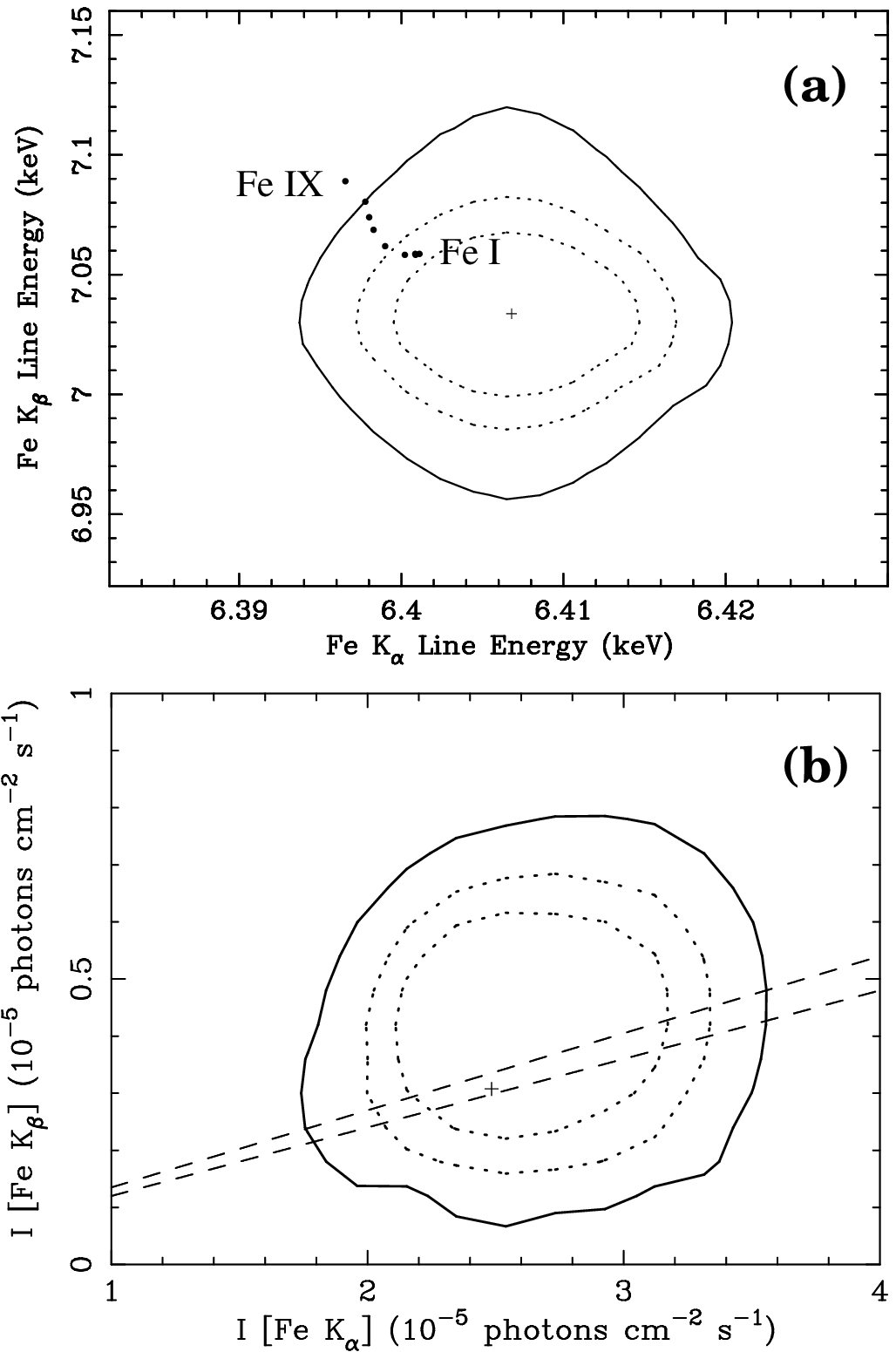}}
  \end{center}
  \caption{(a) The 68\%, 90\%, and 99\% confidence contours
of \ekbeta versus \ekalfap (in the NGC~2992 rest-frame),
measured from the mean XIS spectrum (see section~\ref{fekresults}).
The filled circles are theoretical rest-frame energies of the \fekbeta and \fekalfa
emission
lines from Palmeri \etal~(2003).
(b)
The 68\%, 90\%, and 99\% confidence contours
of \ikbeta versus \ikalfap (in the NGC~2992 rest-frame),
measured from the mean XIS spectrum (see section~\ref{kakbanal}).
The lower and upper solid lines correspond to \fekbetap/\fekalfa branching
ratios of 0.12 and 0.135 respectively (see section~\ref{kakbanal}).
}
\end{figure}

In section~\ref{deconv} and \figidvsin we showed that the intensities of the
narrow and broad \fekalfa emission-line components were decoupled.
The best-fitting line parameters
and statistical errors are shown in \tableresultsp. We obtained only
a lower limit on the disk inclination angle, $\theta_{\rm obs} >31^{\circ}$.
The actual value is somewhat dependent on the model used for the
broad line (here it is the simple Schwarzschild black hole disk
model which does not include light bending -- see Fabian \etal~1989).
However, the overall roughly symmetric shape of the line (\figbroadprofp)
is a characteristic that excludes small inclination angles, independent of the
detailed model, since smaller angles would give rise to an increasing asymmetry
on the red side due to gravitational redshifting.
We obtained an
EW of $118^{+32}_{-61}$~eV for the broad line, which is consistent
with that expected from a ``cold''  accretion disk illuminated by
the observed direct X-ray continuum (e.g. see George \& Fabian 1991).
The index of the line emissivity radial power law, $q = -1.5^{+5.1}_{-0.8}$,
is flatter than the average value for Seyfert~1 galaxies (e.g. Nandra \etal~1997a)
but it is sensitive to the inner and outer disk radius parameters, which we
held fixed for the model-fitting (see section~\ref{baseline}). 
We tried fixing $q$ at the steeper value of $-3$ and obtained a lower limit on 
$\theta_{\rm obs}$ and $R_{\rm in}$ of $29^{\circ}$ and $31r_{g}$
respectively. 

For the centroid energy of the \fekalfa line core
(which we modeled with a Gaussian component), we obtained $6.407 \pm 0.007$~keV
(\tableresultsp). The statistical errors on this centroid energy are
comparable to the residual systematic errors in the absolute 
energy scale of the XIS (on the order of 0.2\% or $\sim 13$~eV at the
Fe~K line energy  --
see section~\ref{meanxis}, Koyama \etal~2006). 
For the corresponding \fekbeta line we obtained a
centroid energy of $7.034^{+0.037}_{-0.038}$~keV. 
The 68\%, 90\%, and 99\% confidence contours of 
the narrow \fekbeta line energy versus the \fekalfa line energy 
are shown in \figekbvsekap,
and \figikbvseka shows corresponding contours for the line intensities of 
\fekalfa and \fekbetap.
{\it Note that both sets of contours were constructed from the baseline
model (section~\ref{baseline}) allowing, in each case,
all thirteen of the remaining parameters to be free}.
The measured energies
and statistical errors on the \fekalfa and \fekbeta line energies are
a function of calibration uncertainties, Doppler and gravitational shifts,
and the ionization state of Fe. Further analysis utilizing
information from the \fekbeta line, including the Fe~K$\beta$/K$\alpha$ ratio,
is described in section~\ref{kakbanal}. 

The measured intensity of the narrow (distant-matter) \fekalfa line,
$I_{N} = 2.5^{+0.7}_{-0.4} \times 10^{-5} \rm \ photons \ cm^{-2} \ s^{-1}$
(see \tableresultsp), is
a factor of $\sim 20$ times larger than the line intensity expected from a
fully-covering Compton-thin spherical shell with the cosmic 
abundance of Fe used in this
paper, the measured line-of-sight column density ($8.0^{+0.6}_{-0.5}
\rm \times 10^{21} \ cm^{-2}$), and the measured intrinsic 2--10~keV luminosity
(see \tableresultsp). Using the expressions in Yaqoob \etal~2001
(see also Krolik \& Kallman 1987), we obtained a predicted
intensity of $1.1 \times 10^{-6} \rm \ photons \ cm^{-2} \ s^{-1}$.
Therefore, if the line emitter is Compton-thin,
one or more of the following must be true:
(i) the Fe abundance is higher than the assumed cosmic value,
(ii) the column density out of the line-of-sight is greater than that 
measured along the line-of-sight, and (iii) there are time delays
between changes in the amplitude of the X-ray continuum luminosity and the
response of the narrow \fekalfa line so that the observed line intensity
may correspond to a time-averaged continuum illumination.
On the other hand, the EW  of $163^{+47}_{-26}$~eV measured for the
narrow \fekalfa line is also compatible with that expected from a line
observed from the X-ray illuminated inner surface of a Compton-thick torus-like
structure (although the historically averaged illuminating luminosity may
need to be higher than the luminosity directly
observed, e.g. see Makishima 1986; Ghisellini, Haardt, \& Matt 1994). 
However, if the \fekalfa line core were produced in
Compton-thick distant matter we would expect to observe
the so-called ``Compton shoulder''
on the red side of the line core. The Compton shoulder would have manifested itself
as an asymmetry in the $E_{\rm Fe \ K\beta}$ versus $E_{\rm Fe \ K\alpha}$ contours
shown in 
\figekbvsekap, in the sense that they would be more elongated towards the
lower energies (e.g. see Yaqoob \etal~2005 and references therein). 
Also, upper limits on the expected Compton-reflection continuum appear
to be too low (section~\ref{refl}). Thus, the distant-matter \fekalfa line
is likely to originate in Compton-thin matter.

\tableresultsc and 
\figewvsfwhm show that the \fekalfa line core (narrow component) is unresolved
by the XIS. We obtained a two-parameter, 99\% confidence upper limit
of $4090 \rm \ km \ s^{-1}$ on the FWHM (or $\sigma_{N} < 37$~eV). 
These upper limits on the line width have been corrected 
(as described in section~\ref{deconv}) for
the fact that the current instrument response matrices do not
model the time-dependent
degradation of the XIS spectral resolution (see Koyama \etal~2006). 
The measured upper limits on the line width can be compared with a
theoretical estimate of FWHM~$\sim 760 \sqrt{(M_{8}/r_{\rm pc})} \rm \ km \ s^{-1}$ 
(assuming a virial relation and an r.m.s. velocity
dispersion of $\sqrt{3} V_{\rm FWHM}/2$; e.g. see Netzer 1990).
Here, $M_{8}$ is the central black-hole mass in units of 
$10^{8} M_{\odot}$, and $r_{\rm pc}$ is the distance (in parsecs), of the line-emitting
structure from the central mass. The mass of the central black hole in
NGC~2992 has been estimated to be $5.2 \times 10^{7} \ M_{\odot}$
from stellar velocity dispersions (Woo \& Urry 2002, and references therein). 
Therefore, the
measured upper limit on the FWHM corresponds to $r_{\rm pc} > 0.018$,
or $r > 21.3$ light days. For comparison, broad ${\rm Pa}\beta$ line emission
has been observed in NGC~2992 with a FWHM of $\sim 1200 \rm \ km \ s^{-1}$
(Rix \etal~1990) indicating $\sim 250$ light days for the location of
the putative hidden BLR (see also Veilleux, Goodrich, \& Hill 1997).

It is pertinent to ask whether the narrow \fekalfa line core has varied
in intensity compared to historical data (see Weaver \etal~1996;
Gilli \etal~2000, and references therein).
Given the large historical amplitude of variability of the continuum 
luminosity (greater than a factor of 20), the response of the narrow
\fekalfa line to the continuum would give some clue about the
size of the line-emitting structure.
In order to answer the line variability question one needs to fit spectra with 
broad and narrow \fekalfa line models simultaneously, but this has not been
done in the literature. We have fitted historical \asca and \bsax data
with the dual Fe~K line model used for the \suzaku data and
have found that at the 99\% confidence level, 
the \asca (1994) and \bsax (1997) data
are consistent with no variability of the narrow \fekalfa line,
and the \bsax (1998) data cannot rule out variability up to a factor of $\sim 5$.
This is a factor of $\sim 4$ less than the continuum variability and
supports an origin of the narrow \fekalfa line core in distant matter. 
At 99\% confidence,
the intensity of the broad
(disk) \fekalfa line during the low-state \asca and \bsax
(1997) observations is consistent with no variability
compared to the \suzaku measurement, but the high state \bsax (1998)
data cannot rule out a factor of up to $\sim 15$ variability.
Therefore the relation between the broad line intensity and continuum
luminosity remains open to future investigation. 
We note that in the 1997 \bsax data (when NGC~2992
was in a near-minimum flux state), there were additional
residuals at $\sim 6.8$~keV (rest-frame) relative to the dual-line model 
possibly due to emission from Fe~{\sc XXV} and/or Fe~{\sc XXVI} 
(see discussion in Gilli \etal~2000).

\subsection{Compton-Reflection Continuum}
\label{refl}

\begin{figure}
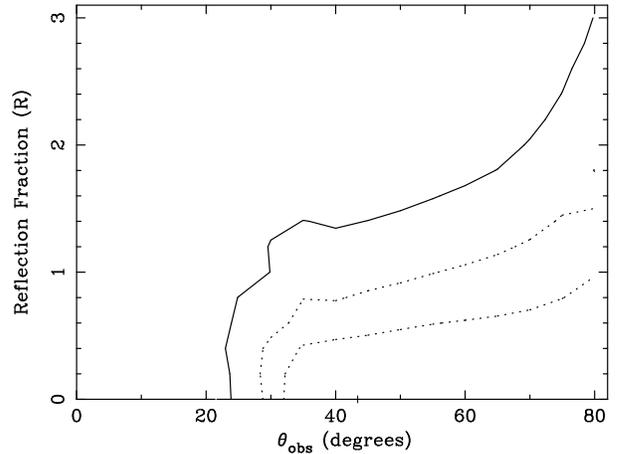

  \begin{center}
    \rotatebox{270}{\FigureFile(60mm,46mm){ty_fig5.ps}}
  \end{center}
  \caption{The 68\%, 90\%, 99\% confidence contours of the relative Compton-reflection
fraction ($R$) versus the inclination angle ($\theta_{\rm obs}$) of the disk normal to the
observer's line-of-sight, based on the best-fitting baseline model (section~\ref{baseline}).
A value of $R=1$ corresponds to a Compton-reflection continuum normalization
expected from a disk subtending a solid angle of $2\pi$ at a central
non-varying X-ray source.
The inclination angle parameters of the disk for the Compton-reflection continuum
and relativistic disk \fekalfa line model components were tied together.
See section~\ref{refl} for details.
}
\end{figure}

If the broad Fe~K line originates in an X-ray illuminated accretion disk
then basic physics predicts a specific Compton-reflection continuum shape
and amplitude commensurate with the Fe~K line parameters (e.g. George \& Fabian
1991; Magdziarz \& Zdziarski 1995) and the
disk ionization state.
We used a simple disk reflection model (see section~\ref{baseline}) to parameterize
the total reflection continuum 
from {\it all} possible locations (e.g. the accretion disk and any 
Compton-thick distant matter) because only a small portion of the
XIS bandpass ($\sim 7-10$~keV) can possibly constrain it. 
The best-fitting value of the so-called reflection fraction, $R$, was 0.
$R$ is the
normalization of the reflection continuum relative to that expected
from a steady-state X-ray illuminated neutral disk subtending a solid angle of
$2\pi$ at the X-ray source. Although the XIS data cannot constrain the
reflection continuum very well, we included it in the model because omitting
it could potentially affect the inferred values of some of the other
model parameters, especially those of the broad disk line.
We found that the upper limits on $R$ were strongly dependent on the
disk inclination angle, principally because of the angle-dependence of
the relativistic line profile shape (the angular dependence of the
reflection continuum shape in the restricted energy band is negligible).
Therefore, we constructed joint 68\%, 90\%, and 99\% confidence contours 
of $R$ versus $\theta_{\rm obs}$, and these are shown in \figrvsthetap.
It can be seen that the 90\% confidence, two-parameter upper limit on
$R$ varies from $\sim 0.6$ to $0.9$ as $\theta_{\rm obs}$ varies from
its lower limit of $\sim 30^{\circ}$ to $\sim 80^{\circ}$ respectively
(the spectral fit is not well-behaved as we approach $\sim 90^{\circ}$).
For comparison with values in the literature,
the 90\%, {\it one-parameter} upper limits on $R$ are 0.60 and 0.78
for fixed inclination angles of
$\theta_{\rm obs}=43^{\circ}$ (the best-fitting value)
and $\theta_{\rm obs}=60^{\circ}$ respectively (see
also \tableresultsp).

We checked that the limits on the Compton-reflection continuum
were not sensitive to background subtraction systematics by
constructing a completely new background spectrum from
rectangular regions from off-source locations in XIS2 and XIS3
and summed over the three observations. Note that the nominal
background spectrum used thus far was constructed from circular
regions centered at different positions and as far as
possible relative to the new background regions (see section~\ref{clnxis}).
With the new background spectrum in place, the baseline model with
fifteen free parameters was re-fitted and there was no statistically
distinguishable change in any of the parameters. The one-parameter,
90\% confidence upper limit on $R$ for $\theta_{\rm obs}=43^{\circ}$
remained at 0.60 and the corresponding upper limit for 
$\theta_{\rm obs}=60^{\circ}$ decreased from 0.78 to 0.76.
Note in \figrvstheta how the data strongly reject models with small
inclination angles ($\theta_{\rm obs}<20^{\circ}$ is ruled out at
$>99\%$ confidence). This is further confirmation that the narrow and
broad Fe~K lines are decoupled because in poorer quality data a
low-inclination angle disk line model can fit the narrow core of the Fe~K line
profile even if the line core has a non-disk origin. 

We also checked that the disk line intensity was not sensitive
to the amplitude of the reflection continuum in the fit by constructing
confidence contours of the disk line intensity, $I_{\rm disk}$,
versus $R$ (and reverting to the nominal background spectrum
for consistency with the earlier analysis). We found that that
the 99\% confidence lower limit on $I_{\rm disk}$ was fairly flat and
non-zero over the range $0<R<3$, and was never less than
$0.4 \times 10^{-5} \ \rm photons \ cm^{-2} \ s^{-1}$. 

Under the stated assumptions of our model
we would expect $R=1$.
The one-parameter upper limits
on $R$ discussed above are not necessarily inconsistent with the 
measured EW ($118^{+32}_{-61}$~eV)
of the broad Fe~K line. The small bandpass leverage may bias $R$, 
and our model for the reflection continuum does not include relativistic smearing
(which would tend to lower the effective value of $R$). The model also
assumes a cold, neutral disk with a cosmic Fe abundance.
However, our upper limits on $R$ 
do imply that there is unlikely to be a significant
Compton-reflection continuum from the structure that produces the
core of the Fe~K fluorescent line emission (e.g. the putative obscuring
torus), which in turn implies that the structure is Compton-thin.

Unfortunately the PDS data from both of the \bsax observations, including the
high-state observation in 1998, were too noisy to obtain useful measurements
or limits on a Compton-reflection continuum (see Gilli \etal~2000).
Although Colbert \etal~(2005) claimed to measure a Compton-reflection continuum
in the off-nuclear region of NGC~2992 using \chandrap, the data only
extended up to 8~keV, the signal-to-noise was poor, and the data suffered from
severe pile-up. The models fitted were complex in relation to the quality
of the data and the uniqueness of the deconvolution of the different
spectral components under such circumstances is an issue that was not addressed.  
In the future we expect the background modeling and subtraction systematics for 
the HXD
onboard \suzaku to improve and we will revisit the challenge of measuring
the Compton-reflection continuum in NGC~2992.
{\it Nevertheless, it is important to note that the detection of the relativistic disk
line in the XIS data is insensitive to uncertainties in the reflection continuum:
there was no preferred model solution
in which a non-zero value of $R$ forced the relativistic disk line intensity to zero.}

\subsection{Upper Limits on the Ni/Fe Abundance}
\label{niabun}

We investigated Ni~$K\alpha$ line emission in the XIS data by adding
another Gaussian line component to the baseline model (section~\ref{baseline}).
The intrinsic width of the additional Gaussian was tied to that of the
\fekalfa and \fekbeta lines. Since visual inspection revealed little
evidence for Ni~$K\alpha$ line emission we fixed the center energy of
the Gaussian at the expected (rest-frame) energy of 7.472~keV (Bearden 1967).
Thus, only the intensity of the Ni~$K\alpha$ line was a free parameter,
giving a total of sixteen free parameters in the model. The results are
shown in \tableresultsp, and it can be seen that a statistically
significant detection was not obtained. The decrease in the fit statistic
was only $\Delta\chi^{2} \sim 1.1$.
We obtained a 90\% confidence, one-parameter upper limit
on the EW of 28~eV. 

If the \fekalfa and Ni~$K\alpha$ lines are formed in Compton-thin matter,
the ratio of their intensities depends principally on the respective ratios of the
fluorescence yields, the $K\beta/K\alpha$ branching ratios, the photoelectric
absorption cross-sections, and the Fe and Ni abundances.
Using Bambynek \etal~(1972) for the fluorescence yields and $K\beta/K\alpha$ branching ratios,
and Henke\footnote{{\tt www-cxro.lbl.gov}}
for photoelectric absorption cross-sections,
we evaluated equation (3) in Yaqoob \etal~(2001) for Fe and Ni and obtained
the theoretical ratio of line intensities:
$I_{\rm Ni \ K\alpha}/I_{\rm Fe \ K\alpha} =  0.96 (A_{Ni}/A_{Fe})$.
From the XIS data we constructed a joint 90\% confidence contour of 
$I_{\rm Ni \ K\alpha}$ versus $I_{\rm Fe \ K\alpha}$ and obtained an upper limit
on $I_{\rm Ni \ K\alpha}/I_{\rm Fe \ K\alpha}$ of 0.176. Thus we obtained
a 90\% confidence upper limit on the Ni/Fe abundance ratio of 0.183. For
comparison, the Ni/Fe abundance ratio from the Anders \& Grevesse (1989)
solar values is 0.038 (i.e. a factor 4.8 smaller than our upper limit).
If the $\rm K\alpha$ lines were observed in reflection from Compton-thick
matter then the expected
$I_{\rm Ni \ K\alpha}/I_{\rm Fe \ K\alpha}$ ratio would be less
than the Compton-thin case (due to absorption of the Ni~$K\alpha$
by Fe) and our upper limit on the Ni/Fe abundance would be larger.

\section{Precision X-ray Spectroscopy with the Fe K${\alpha}$ and Fe K${\beta}$ Lines}
\label{kakbanal}

In this section we will constrain the ionization state of Fe
responsible for the narrow Fe~K line emission in a robust
manner by using as
much information as possible from the spectrum and by utilizing
the latest atomic physics data and associated uncertainties. 
Firstly, we can make an immediate deduction from the very fact that
the \fekbeta fluorescence line is detected at all.
That is, Fe must be less ionized
than Fe~{\sc xvii} in the line-emitting matter responsible for the
narrow Fe~K line emission because Fe~{\sc xvii} and ions with higher ionization
states have no M shell electrons.
We can constrain the ionization state further by taking advantage of the fact
that we measured the \fekalfa and \fekbeta line energies with
sufficiently small statistical errors. We cannot do this
directly from the measured line energies (e.g. using \figekbvsekap) because they
could be affected by residual XIS energy scale uncertainties, possible
gravitational redshifts, and possible Doppler shifts. 
We will describe a method which enables us to place an upper limit
on the dominant ionization state of Fe that is {\it independent
of the residual uncertainty in XIS gain, and does not require
any knowledge of the possible gravitational and Doppler shifts}.
Our result will depend only on the uncertainty in the absolute {\it offset}
in the XIS energy scale, on which we can confidently place
robust limits in general (Koyama \etal~2006),
and for for every observation using the onboard calibration sources. 
However, we first justify
our confidence that the \fekbeta line suffers from negligible contamination
from possible
\feklya line emission.

In \figekbvseka we showed the 68\%, 90\%, and 99\% confidence contours
of \ekbeta versus \ekalfa measured from the XIS data.
The fact that the contours are fairly symmetric about the best-fit
and are not elongated towards lower \fekbeta energies
implies that there is little or no blending from \feklyap, whose
rest-frame centroid energy would be at 6.966~keV (e.g. Pike \etal~1996).
We added another Gaussian line component to the baseline model
(section~\ref{baseline}) at a fixed energy of 6.966~keV, with its intrinsic width
tied to that of the \fekalfa and \fekbeta lines and obtained a best-fit 
\feklya EW of 0 and a 90\%, one-parameter upper limit of 18.5~eV.

The 68\%, 90\%, and 99\% confidence contours of
\ikbeta versus \ikalfa measured from the XIS data are
shown in \figikbvsekap. The 
lower and upper dashed lines
correspond to \ikbetap/\ikalfa ratios of 0.12 and 0.135
respectively. Palmeri \etal~(2003) give a detailed discussion of
the theoretical and experimental values of the \fekbetap/\fekalfa branching ratio.
They compare theoretical values of the ratio in the literature for Fe~{\sc i} to Fe~{\sc ix}
calculated using different codes and assumptions and they compare 
these theoretical values with several experimental measurements for Fe~{\sc i}.
Palmeri \etal~(2003) show that for Fe~{\sc i}, except for one
experimental value, the remainder of the theoretical and
experimental values of the \fekbetap/\fekalfa branching ratio lie
in the range $\sim 0.12-0.135$. Including the highest experimental value,
the range is $\sim 0.12-0.145$ corresponding to a dispersion of $\sim 20\%$.
For higher ionization states of Fe the branching ratio increases, and 
Palmeri \etal~(2003) show that the
theoretical value for Fe~{\sc ix} can be as high as $\sim 0.17$.
The actual value of the observed ratio of the \fekbeta to \fekalfa line
depends a little on the optical depth of the medium in which the lines
originate since the two lines have a slightly different albedo. 

New calculations for the energies of fluorescent \fekalfa
and \fekbeta transitions, fluorescent yields, and rates,
for all charge states of Fe have been
presented by Palmeri \etal~(2003), Mendoza \etal~(2004), Bautista \etal~(2003),
and Kallman \etal~(2004). These works also include a comparison with 
previous calculations (and measurements, where appropriate) in the
literature. In \figekbvsekap, for Fe~{\sc i}--Fe~{\sc ix}, we have plotted the 
expected \fekbeta line energies against the expected \fekalfa line energies
(solid, filled circles), overlaid
on the measured \fekbeta line energy versus \fekalfa line energy confidence
contours. The energies were calculated from
Palmeri \etal~(2003), weighted according to the appropriate branching
ratios for $K\alpha_{1}$ and $K\alpha_{2}$, and for $K\beta_{1}$ and $K\beta_{2}$
where necessary. As discussed in detail by Kallman \etal~(2004),
and references therein, for ionization states higher than Fe~{\sc ix}
the number of transitions increases, and a single line centroid energy
for comparison with observations becomes less and less meaningful since
the fluorescent line profile shape then depends on the 
details of the excitation mechanisms in play.
One must then compute the detailed photoionization equilibrium in order to
derive theoretical line profiles to compare with observations.
From a comparison of results of theoretical calculations with laboratory
measurements (solid state for Fe~{\sc i}, EBIT results for Fe~{\sc ix}),
Kallman \etal~(2004) assess the systematic error in the
Fe~K fluorescent line energies to be $\sim 0.5 \rm \ \rm m\AA$
for Fe~{\sc i} and $\sim 2 \rm \ \rm m\AA$ for Fe~{\sc ix}.
We note that in eV, a systematic error of $\Delta\lambda$ in $\rm m\AA$,
is $\Delta E({\rm eV}) \sim 3.3 \Delta \lambda$ and 
$\Delta E({\rm eV}) \sim 4.0 \Delta \lambda$
at 6.4~keV and 7.0~keV respectively. 

It is important to note from the filled circles in \figekbvseka that
although the theoretical \fekbeta line energy increases along the
sequence Fe~{\sc i} to Fe~{\sc ix}, the \fekalfa line energy
{\it does not increase for these ionization states}. This was pointed out by
Palmeri \etal~(2003) and 
Kallman \etal~(2004) and the results are essentially a refinement of the results
of Makishima (1986) which are universally quoted to describe
a monotonic increase in the \fekalfa line energy with increasing charge
state of Fe. However, Makishima (1986) never gave numerical values for
the \fekalfa line energies, and for the ionization states Fe~{\sc i} to
Fe~{\sc ix} it is not possible to extract the line energy values from their
plot with sufficient precision to compare with the values from Palmeri \etal~(2003).
Although the decrease in the \fekalfa line energy in going from 
Fe~{\sc i} to Fe~{\sc ix} {\it is less than 10~eV, the XIS data for
NGC~2992 are sensitive to these small energy differences}.
The statistical and systematic errors on the \fekalfa line energy 
measured from the NGC~2992 XIS data are comparable to the
difference in the theoretical values of the Fe~{\sc i} and Fe~{\sc ix}
line energies. 
The values of the \fekalfa and \fekbeta line energies,
the associated statistical errors, and confidence contours
measured for NGC~2992 
are not only a function of the 
predominant charge state of Fe and residual calibration uncertainties,
but are
also affected by possible Doppler and gravitational shifts.
In order to make inferences on the
ionization state of Fe from the XIS data we must consider all of these
factors, including uncertainties in the
theoretical line energies, so {\it we cannot simply deduce the ionization states
allowed by the data from \figekbvsekap}.

In order to constrain the dominant ionization stage of Fe further,
we can recast the \ekbeta versus \ekalfa  contours shown in \figekbvsekap.
For a given ionization stage of Fe there are a pair of ``true''
(or theoretical) values
of \ekalfa and \ekbetap.
Suppose that the observed and true values
of \ekalfa and \ekbeta  are linearly related by

\begin{equation}
E_{{i, {\rm Fe \ K}\alpha}} \ ({\rm observed})  
=  A + B \times  E_{{i, {\rm Fe \ K}\alpha}} \ ({\rm true}) 
\end{equation}

\par\noindent
and

\begin{equation}
E_{{i, {\rm Fe \ K}\beta}} \ ({\rm observed})  
=  A + B \times  E_{{i, {\rm Fe \ K}\beta}} \ ({\rm true}),  
\end{equation}

\par\noindent
where $i = 0$ to 25 represents the ionization state of Fe
(i.e. the number of electrons missing in the ion relative to
the neutral atom),
$A$ represents any residual calibration uncertainties in the 
{\it offset} 
of the absolute energy scale and $B$ includes contributions from residual
uncertainties in the XIS gain as well as possible gravitational and/or
Doppler shifts. Certainly, over the restricted energy range
being considered here, any higher order terms in the gain uncertainty
are negligible (see Koyama \etal~2006), and the gravitational
and Doppler terms only make a linear contribution. Therefore, the
linear equations above are an excellent representation of the relations
between the true rest-frame line energies and the final line energies
measured by the XIS.

\begin{figure}
  \begin{center}
    \rotatebox{0}{\FigureFile(80mm,61mm){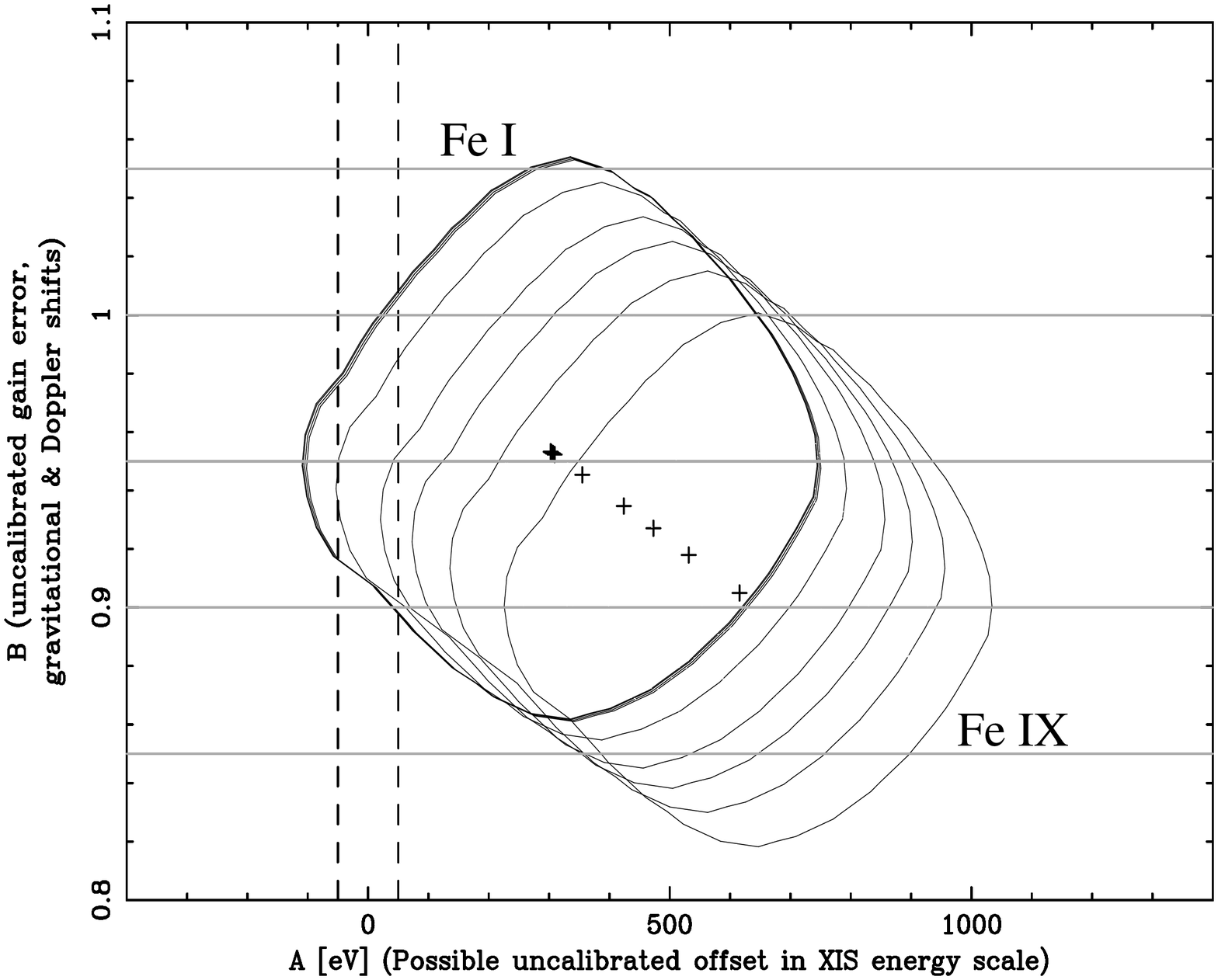}}
  \end{center}
  \caption{The 99\% confidence contours of $B$ versus $A$ calculated from the
99\% contour of \ekbeta versus \ekalfa in \figekbvseka (which was derived
directly from the XIS spectrum of NGC~2992),
for Fe~{\sc i} to Fe~{\sc ix}.
The method of transformation of the \ekbeta versus \ekalfa contour
is described in section~\ref{kakbanal}.
The \fekalfa and \fekbeta fluorescent line
energies were calculated using atomic data in
Palmeri \etal~(2003) and are also shown
as filled circles in \figekbvsekap.
The variable $A$ represents residual uncertainties in
the offset of the absolute energy scale which is believed to be on
the order of $\sim 13$~eV or less, but the dashed vertical lines are placed very
conservatively at $\pm 50$~eV. $B$ represents the product of any gravitational
shift, Doppler shift, or residual uncorrected gain factors. Contours of
$B$ versus $A$ that do not have any portion lying inside $A = \pm 50$~eV are
ruled out at 99\% confidence.
}
\end{figure}

Since $A$ and $B$ are the same for the
Fe~K$\alpha$ and Fe~K$\beta$ lines, 
we can solve for $A$ and $B$ for a given ionization stage of Fe
for any given pair of measurements of $E_{{i, {\rm Fe \ K}\alpha}}~({\rm observed})$
and $E_{i, \rm Fe \ K\beta}~({\rm observed})$.  Thus
we can recast a contour of the measured energies \ekbeta and \ekalfa 
for a given confidence level 
into a contour of $B$ versus $A$ for every $i$ (i.e. we can generate up
to 25 contours from \figekbvsekap). 
Since we have independent information on the calibration of the
absolute energy scale of the XIS data,
we can then rule out any ionization states
that yield $B$ versus $A$ contours that give solutions for $A$ that
are inconsistent with the known accuracy of the XIS energy scale.
The absolute energy scale calibration is good to 0.2\% at the \fekalfa line energy
and $\sim \pm 5$~eV below 1~keV (Koyama \etal~2006). 

The 99\% confidence contours of $B$ versus $A$ calculated as described
above are shown in \figfeion for Fe~{\sc i} to Fe~{\sc ix}, 
using the expected \fekalfa and \fekbeta line
energies calculated from Palmeri \etal~(2003).
These energies were also shown as filled circles in \figekbvsekap.
The vertical dashed lines in \figfeion represent the uncertainty
in the absolute energy scale offset of the XIS (i.e.
they enclose the possible range of values of $A$). Although we have stated that this
is nominally $\sim \pm 13$~eV or less, we will base our conclusions on the {\it very
conservative} assumption that $|A|$ is less than $50$~eV. The 
vertical dashed lines are therefore at $A = \pm 50$~eV. It can be
seen from \figfeion that the contours for Fe~{\sc vii} to Fe~{\sc ix}
are ruled out {\it regardless of the value of B}. 
We can go further and take into account the 
uncertainties in the theoretical values of the \fekalfa and \fekbeta
line energies. The $B$ versus $A$ contours depend on \ekbetap$-$\ekalfa
and we considered the two extreme cases, corresponding to the theoretical difference
in these energies hypothetically
being in error by $-2 \ \rm m\AA$ and $+2 \ \rm m\AA$. We recalculated
the $B$ versus $A$ confidence contours for each of these cases for
Fe~{\sc i} to Fe~{\sc ix}. 
The cases for a theoretical error of $-2 \ \rm m\AA$ and $+2 \ \rm m\AA$ 
gave contours which ruled out ionization states higher than Fe~{\sc v}
and Fe~{\sc vii} respectively. 
Therefore, we can say that at 99\% confidence the narrow 
core of the \fekalfa line in NGC~2992 originates in matter in which
the predominant ionization state of Fe lies in the range Fe~{\sc i}--Fe~{\sc vii}.
{\it We emphasize that 
this conclusion requires no knowledge of the uncertainty in the
XIS gain, or of any possible gravitational or Doppler shifts experienced
by the photons leaving the source}.

\section{Summary and Conclusions}
\label{concl}

We have presented detailed X-ray spectroscopy in the 0.5--10~keV 
band of the Seyfert~1.9 galaxy NGC~2992 with the \suzaku XIS. Here we summarize
the results.

\begin{enumerate}

\item The mean observed 2--10~keV flux of NGC~2992 during the three \suzaku
observations in 2005 November and December was 
$\sim 1.2 \times 10^{-11} \rm \ erg \ cm^{-2} \ s^{-1}$, a factor of
$\sim 3$ higher than the historic minimum flux, and a factor of $\sim 6$
lower than the historic maximum. The absorption-corrected intrinsic
2--10~keV luminosity was $1.6 \times 10^{42} \rm \ erg \ s^{-1}$. 

\item We detected the extended, optically-thin soft X-ray thermal emission 
found with previous X-ray astronomy missions, and
obtained a
temperature of $kT=0.656^{+0.088}_{-0.061}$~keV. This is the most precise
measurement of $kT$ for NGC~2992 to date although the errors are
purely statistical and do not include any possible systematic effects. 
We obtained a luminosity for the soft emission component 
of $\sim 1.2 \pm 0.4 \times 10^{40} \rm \ erg \ s^{-1}$ which is consistent
with that deduced from \chandra data (Colbert \etal~2005).

\item We modeled the intrinsic X-ray continuum with a power law and
obtained $\Gamma = 1.57^{+0.06}_{-0.03}$ for the power-law
photon index.

\item We derived a column density of $8.0^{+0.6}_{-0.5} \times 10^{21} \rm \ cm^{-2}$
for the line-of-sight X-ray obscuration,
which corresponds to a
Thomson depth of $\sim 0.006^{+0.005}_{-0.004}$ (i.e. the absorber
is Compton-thin, at least in the line-of-sight). This column density is
consistent with values obtained from previous \bsax and \asca observations.

\item We estimated $\tau_{\rm es} = 0.10 \pm 0.03$
for the optical depth of the extended, warm, electron scattering zone
which is responsible for scattering $(7.3 \pm 2.1)\%$ of the intrinsic
hard X-ray continuum into the line-of-sight.

\item We detected broad and narrow Fe~K line emission. When we modeled the
broad and narrow lines with line emission from a 
relativistic disk around a Schwarzschild black hole and a Gaussian
profile respectively, we found that the intensities of the broad and
narrow lines were decoupled at a confidence level of greater than $3\sigma$.
We obtained an EW of $118^{+32}_{-61}$~eV, a radial line-emissivity law
of $r^{-1.5 (-0.8,+5.1)}$, and a
lower limit of $31^{\circ}$ on the disk inclination angle
from the disk-line model. The actual values are model-dependent, however.

\item The narrow Fe~K line core EW was $163^{+47}_{-26}$~eV. 
The line core 
was unresolved by the XIS and we obtained a 99\% confidence, two-parameter
upper limit on the FWHM of $4090 \rm \ km \ s^{-1}$,
after taking into account the time-dependent degradation in the
XIS spectral resolution (as discussed in section~\ref{meanxis} and section~\ref{deconv}). 

\item The absolute flux
in the narrow line is a factor of more than $20$ larger than that
expected from a
fully-covering spherical shell 
illuminated by the measured intrinsic 2--10~keV luminosity,
having a cosmic abundance of Fe and 
a uniform column density equal to
that measured in the line-of-sight.
Therefore, roughly speaking, the product of the solid angle subtended
by the line-emitting structure at the source, its angle-averaged column density,
its Fe abundance, and the mean historical 2--10~keV luminosity averaged
over timescales of the order of the structure light-crossing time, must be $>20$.

\item We detected the \fekbeta fluorescent line corresponding to the 
\fekalfa line core. Taking into account residual uncertainties in 
the XIS energy scale calibration and conservative uncertainties 
on the theoretical energies of the \fekalfa and \fekbeta transitions for
different ionic species of Fe, we were able to constrain the predominant
ionization state of Fe responsible for producing the narrow Fe~K line core
to be Fe~{\sc i}--Fe~{\sc vii}. Ionization states of Fe~{\sc viii} 
or higher are
ruled out at 99\% confidence. This is the most precise determination to date of
the ionization state of the matter emitting the Fe~K line core
and was achieved using a method
that {\it requires no knowledge
of any uncertainties in the XIS gain and any
possible gravitational and Doppler shifts}.

\item We were able to decouple the disk and distant-matter
\fekalfa emission lines because the continuum flux was low, giving a 
large EW for the line core, with good signal-to-noise ratio. The large core EW also
enabled the \fekbeta line to be measured with a good signal-to-noise ratio,
giving redundant information on the width of the distant-matter lines.
This demonstrates the importance of measuring the distant-matter
Fe~K line emission parameters in order to measure 
those of the relativistic Fe~K line.

\item Although the XIS data do not have the bandpass to measure the
Compton-reflection continuum associated with the broad Fe~K line,
we included it in all of the spectral analysis with a parameterized
normalization, $R$, relative to that expected from a steady-state
illumination of a disk subtending a solid angle of $2\pi$ at the X-ray source.
We obtained upper limits on $R$ as a function of
the disk inclination angle (constrained by the relativistic Fe~K line). 
We found (at 90\% confidence for two parameters),
$R<0.9$ for an inclination angle of  $80^{\circ}$, going down to
$R<0.6$ when the inclination angle is at its lower limit of $31^{\circ}$. 
If all of this reflection is associated with the broad Fe~K line,
which presumably originates in a Compton-thick disk, there is
no room left for reflection from distant matter.
This implies that
the narrow Fe~K line core is likely to originate in Compton-thin matter,
supported by the lack of a Compton shoulder on the Fe~K line core.
Although the EW of the Fe~K line core is too large for steady-state
illumination of Compton-thin matter, several possible factors
can account for it (see [8] above).

\end{enumerate}

In the future we will refine our modeling by utilizing the \suzaku HXD data
when the background model has achieved the goal of 1\% systematic uncertainties.
The HXD data will better constrain the Compton-reflection continuum components.
We will also improve model constraints by including data from XIS0 and XIS1
when the calibration has advanced further.
Future observations with higher spectral resolution, such as
gratings or calorimeters, will be able to resolve the narrow core of the \fekalfa
line and thereby better constrain the location of the line-emitter. In addition, improved
throughput will enable better measurement of the profile of the broad Fe~K line.
Further monitoring of NGC~2992  to measure
variability in the Fe~K line components and in the reflection continuum as the continuum
undergoes large excursions in amplitude promise to
reveal important clues on the accretion
process and on the structure of the central engine beyond the accretion disk.

%\section{Acknowledgments}
We thank the \suzaku team and operations staff for their hard work and dedication 
that has made these observations of NGC~2992 possible and that has
facilitated the calibration and analysis of the data. We also thank
T. Kallman, T. Dotani, M. Bautz, K. Koyama, and H. Kunieda
for helpful discussions and comments.
TY, KM, AP, and JR
acknowledge the support of NASA grant NNG04GB78A, a cooperative agreement
between the Johns Hopkins University and the NASA/Goddard Space Flight Center. 
This research
made use of the HEASARC online data archive services, supported
by NASA/GSFC. This research also made use of the NASA/IPAC Extragalactic Database
(NED) which is operated by the Jet Propulsion Laboratory, California Institute
of Technology, under contract with NASA.

%%%%%%%%%%%%%%%%%%%%%%%%%%%%%%%%%%%%%%%%%%%%%%%%%%%%%%%%%%%%%%%%%%%%%

%%%%%%%%%%%%%%%%%%%%%%%%%%%%%%%%%%%%%%%%%%%%%%%%%%%%%%%%%%%%%%%%%%%%%%%%%%%%%%%%%

\end{document}